\documentclass{aa}  

\usepackage{graphicx}
\usepackage[varg]{txfonts}
\usepackage{lscape}
\usepackage{natbib}
\usepackage{rotating}

\def\deg{\ifmmode^\circ\else$^\circ$\fi}
\def\alphaTF{\ifmmode{\alpha_{\mathrm{\,{\small TF}}}}\else{$\alpha_{\mathrm{\,{\small TF}}}$}\fi}

\begin{document}

\title{Evolution along the sequence of S0 Hubble types\\induced by dry minor mergers.}
\subtitle{II - Bulge-disk coupling in the photometric relations through\\merger-induced internal secular evolution}

\author{M.~Carmen Eliche-Moral\inst{1}, A.~C\'esar Gonz\'{a}lez-Garc\'{\i}a\inst{2,3,4}, J.~Alfonso L.~Aguerri\inst{3,4}, Jes\'us Gallego\inst{1}, \\Jaime Zamorano\inst{1}, Marc Balcells\inst{3,4,5}, \and Mercedes Prieto\inst{3,4}
            }

\institute{Departamento de Astrof\'{\i}sica y CC.~ de la Atm\'osfera, Universidad Complutense de Madrid, E-28040 Madrid, Spain\\ \email{mceliche@fis.ucm.es}
  \and
Present address: Instituto de Ciencias del Patrimonio, CSIC, R\'{u}a San Roque 2, Santiago de Compostela, E-15704 A Coru\~{n}a,  Spain \and 
Instituto de Astrof\'{\i}sica de Canarias, C/ V\'{\i}a L\'actea, E-38200 La Laguna, Tenerife, Spain
  \and
Departamento de Astrof\'{\i}sica, Universidad de La Laguna, E-38200 La Laguna, Tenerife, Spain
  \and
Isaac Newton Group of Telescopes, Apartado 321, E-38700 Santa Cruz de La Palma, Canary Islands, Spain
}

   \date{Received 2012 December 3, 2012; accepted 2013 February 5}

\abstract{
Galaxy mergers are widely discussed as one possible evolution mechanism for lenticular galaxies (S0s), because even minor mergers induce structural changes that are difficult to reconcile with the strong bulge-disk coupling observed in the photometric scaling relations of these galaxies.
}{
We check if the evolution induced onto S0s by dry intermediate and minor mergers can reproduce the S0 photometric scaling relations.
}{
We analyse the bulge-disk decompositions of the collisionless N-body simulations of intermediate and minor mergers onto S0s presented previously to determine the evolution induced by the mergers in several relevant photometric planes.
}{
The mergers induce an evolution in the photometric planes that is compatible with the data of S0s, even in the relations that indicate a strong bulge-disk coupling. Mergers drive the formation of the observed photometric relation in some cases and induce a slight dispersion compatible with data in others. Therefore, this evolutionary mechanism tends to preserve the observational photometric relations. In the photometric planes where the morphological types segregate, the mergers always induce evolution towards the region populated by S0s. No clear trends with the mass ratio of the encounter, the central satellite density, or the spin-orbit coupling are found for the range of values studied. Long-pericentre orbits generate more concentrated disks and expanded bulges than initially, while short-pericentre orbits do the opposite. The structural coupling of the bulge and the disk is preserved or reinforced in the models because mergers trigger internal secular processes in the primary disk that induce significant bulge growth. This happens even though the encounters do not induce bars in the disks.
}{
Intermediate and minor mergers can be considered to be plausible mechanisms for the evolution of S0s if one includes their photometric scaling relations, because they can preserve and even strengthen any pre-existing structural bulge-disk coupling by triggering significant internal secular evolution, even without bars or dissipational effects. Satellite accretions thus seem to unavoidably entail internal secular evolution, meaning that it may be quite complex to isolate the effects of the internal secular evolution driven by mergers from the one due to purely intrinsic disk instabilities in individual early-type disks at the present.
}


\titlerunning{Evolution along the sequence of S0 Hubble types through dry minor mergers. II}
\authorrunning{Eliche-Moral et al.}

   \maketitle

\section{Introduction}
\label{Sec:introduction}

According to the current hierarchical models of galaxy formation, galaxy disks are formed first through gas cooling in collapsing dark matter halos \citep{2003Ap&SS.284..397D,2009ApJ...705L.133M}. Relevant spheroidal galaxy components appear later in the cosmic scenario, essentially formed through major and/or minor mergers \citep{1999MNRAS.310.1087S}. The later accretion of the left-over gas and stripped stars onto the surviving original disk can rebuild it. In particular, ellipticals would result from extreme merger events with no disk re-building \citep[see, e.g.,][]{2006MNRAS.366..499D}. This scenario is becoming more complex as cosmological simulations improve, indicating that present-day bulges must have been built up in two phases \citep[a fast early collapse at $z>1.5$, followed by a smoother phase driven by mergers or disk instabilities, see][]{2013ApJ...763...26O}. These processes would generate the whole sequence of spiral Hubble types, in such a way that the galaxies that have undergone the richest merger histories host the largest bulges at present \citep{2006ApJ...647..763M,2008MNRAS.387..364Z,2013MNRAS.428..718O}.

Within this cosmological scenario, the bulges are expected to exhibit photometric and kinematical properties similar to the ellipticals, because their formation mechanisms are similar. Moreover, the structures of the bulge and disk in a galaxy should be more clearly decoupled with the rising contribution of merging to the buildup of the galaxy, because the survival of the original disk is less probable in violent or successive mergers \citep{2009ApJ...691.1168H}. This is even more probable if the merger had taken place at $z> 1.5$, when galaxies hosted high fractions of cold, turbulent, and clumpy gas \citep{2011ApJ...730....4B}. The bulge and disk components are thus expected to result from more independent and non-contemporary mechanisms as we consider earlier Hubble types \citep{2009MNRAS.396..696S}, meaning that the bulges and disks of S0 galaxies should be the most clearly decoupled of all types. 

However, recent observations are in conflict with this evolutionary scenario. \citet[L10 hereafter]{2010MNRAS.405.1089L} reported strong photometric scaling relations between the bulges and disks of nearby S0s, suggesting that the formation processes of their structural components must have been coupled. These authors also found that the photometric properties of their bulges are more similar to the bulges of spirals than to ellipticals \citep[see also][]{2008MNRAS.388.1708G,2009ApJS..182..216K,2012ApJS..198....2K}. Moreover, \citet{2012MNRAS.tmp..278K} reported that the (Sersic$+$exponential) photometric decomposition of a sample of 180 unbarred galaxies from the ATLAS$^\textrm{3D}$ survey 
is only a rough approximation to the real dynamical and structural properties of these galaxies (mostly S0s). This suggests that photometric decompositions do not clearly identify decoupled disk- and spheroid-type components that would have formed at different epochs. All these results seem to imply that the contribution of merging to the buildup of present-day S0s must have been negligible \citep[][]{2009ApJ...692L..34L,2010MNRAS.405.1089L}. Other processes that are apparently more compatible with this bulge-disk connection, are thus being proposed to explain the formation of S0s, such as ram-pressure stripping and internal secular evolution \citep[see the review by][]{2012AdAst2012E..28A}.

Nevertheless, the majority of S0s at low redshifts ($\sim 60$\%) reside in groups, an environment where tidal interactions and mergers are observed to drive galaxy evolution \citep[][and references therein]{1982ApJ...257..423H,2006ApJS..167....1B,2007ApJ...655..790C,2009ApJ...692..298W,2012MNRAS.427..790S}. Indeed, many observational studies indicate that mergers and galaxy interactions may have contributed noticeably to the evolution of S0s \citep[][]{2001AJ....121..140K,2004AJ....127.1371K,2006ApJ...639..644E,2009AJ....138..579K,2009A&A...504..389C,2009AJ....138..295F,2009MNRAS.394.1713K,2011MNRAS.411.2148K,2009ApJ...694.1550S,2010AJ....140.1462S,2010ApJ...713..637B,2010A&A...519A..55E,2010arXiv1003.0686E,2011A&A...525A..20C,2011MNRAS.417..863D,2012ApJ...753...43K,2013MNRAS.428..999P}. Independently of whether mergers have partially shaped the characteristics of the S0 population or have built them, the strong bulge-disk coupling observed in these galaxies is difficult to reconcile with the popular view that mergers (even minor ones) efficiently decouple both components \citep[][L10]{2009ApJ...692L..34L}. 

But do mergers unavoidably destroy any pre-existing structural link between the bulge and the disk in a galaxy? Merger simulations confirm this idea for major mergers in general (i.e., in galaxy interactions with mass ratios below 4:1), because they typically result in elliptical-like remnants \citep[][]{2003ApJ...597..893N,2005A&A...437...69B,2005MNRAS.357..753G,2006MNRAS.372L..78G,2006ApJ...636L..81N,2006MNRAS.369..625N,2007ApJ...658...60B}. However, simulations also show that S0-like remnants can result from mergers of intermediate mass ratios \citep[between 4:1 and 7:1, see][]{1998ApJ...502L.133B,2005A&A...437...69B,2006A&A...457...91E,2011A&A...533A.104E}, and from some particular cases of major mergers  \citep{2005ApJ...622L...9S,2007MNRAS.374.1479G,2009MNRAS.398..312G,2009ApJ...691.1168H}. Additionally, minor mergers (i.e., mass ratios above 7:1) onto pre-existing S0s also evolve the galaxy within the sequence of S0 Hubble types \citep{2012A&A...547A..48E}. Nevertheless, it has not been investigated so far whether these S0 remnants reproduce the strong bulge-disk coupling observed in real S0s or not.

In the first paper of this series  \citep[Paper I henceforth]{2012A&A...547A..48E}, we reported that dry intermediate and minor mergers can produce a significant bulge growth in an S0 galaxy, but affect only very little the scale-lengths of its bulge and disk at the same time. Indeed, the evolution induced by the mergers is compatible with the observational distribution of S0s in the space $n$-$r_\mathrm{eff}$-$h_\mathrm{D}$-$r_\mathrm{eff}/h_\mathrm{D}$-$B/T$, where $n$ is the Sersic index, $r_\mathrm{eff}$ the bulge effective radius, $h_\mathrm{D}$ the disk scale-length, and $B/T$ the bulge-to-total luminosity ratio. In the present paper, we complete the structural analysis started in Paper I, by presenting a detailed comparison of the evolution induced by these mergers in the most relevant photometric planes and scaling relations with data of real S0s. We show that the traditional picture of mergers as events that decouple bulges from disks in galaxies is questionable, thus reconciling the merger-related evolution of many S0s with their observed bulge-disk coupling.

The paper is organised as follows. We describe the models and the bulge-disk decompositions performed on the remnants in Sects.~\ref{Sec:models} and \ref{Sec:photometricParameters}. Section\,\ref{Sec:scaling} is devoted to the model scaling. The results are presented in Sect.~\ref{Sec:results}. In Sect.~\ref{Sec:bulgeall}, we analyse the photometric relations including only bulge parameters. The photometric scaling relations involving only disk parameters are discussed in Sect.~\ref{Sec:disk}, and we discuss the photometric planes relating bulge and disk parameters in Sect.~\ref{Sec:bulgedisk}. Finally, the discussion and the main conclusions are given in Sects.~\ref{Sec:discussion} and \ref{Sec:conclusions}. A concordant cosmology is assumed throughout the paper \citep[$\Omega_\mathrm{M} =0.3$, $\Omega_\Lambda =0.7$, $H_0 = 70$\,km s$^{-1}$ Mpc$^{-1}$,][]{2007ApJS..170..377S}. All magnitudes are in the Vega system.

\section{Model description} 
\label{Sec:models}

We used the battery of collision-less simulations of intermediate and minor mergers described in \citet[EM06 henceforth]{2006A&A...457...91E} and in \citet[EM11 hereafter]{2011A&A...533A.104E}, which are 16 experiments in total. Both the primary galaxy and the satellite consist of a bulge-disk-halo structure. All initial models represent S0 galaxies (i.e., with no spiral arms in the disk, and without gas or star formation) with different bulge-to-disk mass ratios ($B/D$). The experiments differ in the orbits, the spin-orbit coupling, the satellite-to-primary mass and density ratios, and the $B/D$ of the initial primary galaxy (S0b or S0c). These models have been extensively described in EM06, EM11, and in Paper I, therefore we only briefly describe them here. 

The majority of the experiments were run using an S0b primary galaxy ($B/D=0.5$), using 185k particles in total for the simulation. Two experiments have an S0c primary galaxy ($B/D=0.08$), using 415k particles in total for the simulation. The total mass-to-light ratios are close to observations in the two galaxy models if an adequate $M_\mathrm{lum}/L$ ratio is assumed ($M/L\sim 10$, where $M = M_\mathrm{lum}+M_\mathrm{dark}$). Satellites are scaled replicas of the primary galaxy with a big bulge in all experiments. The size-mass scaling of the galaxies was set in each experiment by forcing both the primary and satellite galaxies to obey the Tully-Fisher relation: $L \sim V^\alphaTF$, where $L$ is the galaxy luminosity and $V$ its rotational velocity \citep{1977A&A....54..661T}. Several values of the exponent in this relation were considered, spanning the observational range ($\alphaTF=3.0$, 3.5, and 4.0). A higher $\alphaTF$ exponent results in a denser satellite compared to the primary galaxy. 

The galaxies were modelled using the {\tt GalactICS} code \citep{1995MNRAS.277.1341K} and have a bulge$+$disk$+$halo structure. The dark halo is built following an Evans profile, which is modified accordingly in the models with a small bulge to ensure disk stability, and in that cases it may be considered cuspy. Each galaxy was relaxed for about ten disk dynamical times prior to the merger simulation, to guarantee that the disks do not develop relevant internal secular evolution in isolation. The Toomre Q parameter is close to 1 and flat for a large fraction of the disk for all models prior to relaxation. After relaxation, the chosen disk velocity dispersion heats the disk with Q = 1.7, enough to prevent bar instabilities in the disk in isolation. We measured the thickening of the disks by evolving the initial primary galaxy model in isolation for 100 time units, the typical duration of the experiments. The disk scale-height at $h_\mathrm{D}$ typically increases by $\sim 50$\% due to two-body effects, growing by less than $\sim 1/10$ of the disk scale-length at most (EM06). Therefore, the disk heats but is not greatly puffed-up. 

The orbits are parabolic with an initial separation equal to 15 times the primary disk scale-length in all  simulations. We ran experiments with a pericentre distance equal to $h_\mathrm{D}$ and to $8h_\mathrm{D}$. Direct orbits have an initial inclination between the orbital plane and the primary galaxy plane equal to 30\deg. Retrograde orbits use an inclination angle equal to 150\deg. The evolution of each experiment was computed using the {\small TREECODE\/} \citep{1987ApJS...64..715H,1989ApJS...70..419H,1990ApJ...356..359H,1990ApJ...349..562H} and the GADGET-2 codes \citep{2001NewA....6...79S,2005MNRAS.364.1105S}. We checked that the total energy is preserved to better than 0.1\%. Forces were computed to within 1\% of those given by a direct summation for the chosen execution parameters (see EM06 and EM11). All models are evolved for $\sim$2-4 halo crossing times beyond a full merger to ensure that the remnant is relaxed enough (see full-merger and total computed times for each experiment in Table\,2 in EM06 and Table\,3 in EM11). 

Table\,\ref{tab:models} lists the main characteristics of each merger experiment. We ran mergers with satellite-to-primary stellar mass ratios equal to 6:1, 9:1, and 18:1. We refer to each model throughout the paper according to the code used in Paper I: M$m$P[l/s][D/R][b/s], where $m$ refers to the mass ratio of the encounter ($m=6$, 9, or 18 for mass ratios equal to 6:1, 9:1, 18:1, respectively), "Pl" indicates long pericentre ($d_\mathrm{per} = 8h_\mathrm{D}$) and "Ps" short pericentre ($d_\mathrm{per} = h_\mathrm{D}$), "D" is used in direct mergers and "R" in retrograde ones, and the next "b" or "s" letter indicates if the primary galaxy had a big or small bulge ($B/D=0.5$ or 0.08). A final "TF3" or "TF4" suffix is used to refer to the models that assume $\alphaTF=3.0$ or 4.0. Otherwise, a scaling with $\alphaTF=3.5$ must be assumed. For more details, we refer to EM06, EM11, and Paper I.

\begin{table}
\caption{Orbital and scaling parameters of each merger experiment}
\label{tab:models}
{\small
\setlength{\tabcolsep}{0.6em}
\begin{tabular}{l@{\hskip 0.2em}l@{\hskip 0.2em}l@{\hskip 0.2em}l@{\hskip 0.2em}lccrcc}
\hline\hline
\multicolumn{5}{c}{Model code} &   $M_{\mathrm{1}}/M_{\mathrm{2}}$  & $d_{\mathrm{per}}/h_\mathrm{D}$  & \multicolumn{1}{c}{$\theta _1$ ($^o$)} &  \multicolumn{1}{c}{\textrm{$(B/D)_1$}} & $\alphaTF$ \\
\multicolumn{5}{c}{(1)}    & \multicolumn{1}{c}{(2)} & \multicolumn{1}{c}{(3)}         & \multicolumn{1}{c}{(4)}    & \multicolumn{1}{c}{(5)}        & \multicolumn{1}{c}{(6)}       \vspace{0.05cm}\\\hline\vspace{-0.3cm}\\
 (a) & M6 & Ps& Db &      &  6:1   &  0.73 &  30 & 0.5   & 3.5 \\
 (a2)& M6 &Ps& Db& TF3    &  6:1   &  0.73  &  30  & 0.5  & 3.0 \\
 (a3)& M6 & Ps& Db& TF4    &  6:1   &  0.73  &  30  &0.5  & 4.0 \\
 (b)& M6 &Ps &Rb &           &  6:1   &  0.73  & 150 & 0.5  & 3.5  \\
 (c) &M6 &Pl &Db&    &  6:1 &  8.25 &  30    & 0.5   & 3.5 \\
 (d) &M6& Pl& Rb&   &  6:1  &  8.25  &  150  & 0.5   & 3.5 \\
 (e) &M6& Ps& Ds&   &  6:1 &  0.87  &  30    & 0.08  & 3.5 \\
 (f) &M6& Ps& Rs&   &  6:1  &  0.87  & 150   & 0.08  & 3.5\\ \vspace{-0.4cm}\\\hline\vspace{-0.3cm}\\
 (g)& M9& Ps& Db&  & 9:1 &  0.79  &  30   &  0.5   & 3.5 \\
 (g2)& M9& Ps& Db& TF3  & 9:1 &  0.79  &  30   &  0.5  & 3.0 \\
 (g3)& M9& Ps& Db& TF4  & 9:1 &  0.79 &  30   & 0.5  & 4.0 \\
 (h) &M9& Ps& Rb&          & 9:1  &  0.79  & 150  & 0.5   & 3.5\\ \vspace{-0.4cm}\\\hline\vspace{-0.3cm}\\
 (i)& M18& Ps& Db&  & 18:1 &  0.86 &  30    & 0.5  & 3.5 \\
 (j)& M18& Ps& Rb&     & 18:1 &  0.86  & 150   & 0.5   & 3.5 \\
 (k)& M18& Pl& Db& &  18:1 & 8.19  &  30  & 0.5   & 3.5\\
 (l)& M18& Pl& Rb&  &  18:1 & 8.19 &  150  & 0.5  & 3.5 \\\hline\\
\end{tabular}
\tablefoot{\emph{Columns}: (1) Model code: M$m$P[l/s][D/R][b/s][TF3/4], see text. (2) Luminous mass ratio between the primary galaxy and the satellite. (3) First pericentre distance of the orbit in units of the original primary disk scale-length. (4) Initial angle between the orbital momentum and the primary disk spin (direct and retrograde orbits). (5) Bulge-to-disk ratio of the original primary galaxy used in the experiment: $B/D=0.5$ (big bulge) or $B/D=0.08$ (small bulge). (6) Value of $\alphaTF$ assumed for the satellite scaling to the primary galaxy. } 
}
\end{table}

\section{Bulge-disk photometric decompositions of the remnants} 
\label{Sec:photometricParameters}

Our initial primary galaxy models could be equated with late-type S0s (S0b and S0c) in the scheme proposed by \citet{1976ApJ...206..883V} and recently updated by \citet{2012ApJS..198....2K} \citep[see also L10;][]{2011MNRAS.416.1680C}, especially due to their lack of gas and spiral patterns. In paper I, we show that these mergers induce noticeable bulge growth and that the final remnants are also S0s with higher $B/T$ values. Therefore, the mergers induce an evolution in the S0 sequence of Hubble types towards earlier types (in particular, S0c$\longrightarrow$S0b and S0b$\longrightarrow$S0a). 

The remnants of these experiments do not have significant non-axisymmetric components or distortions (EM06; EM11). Considering this, we derived azimuthally averaged 1D surface mass density profiles of their stellar material, using face-on views to compare them with observations, which usually correct for galaxy inclination. These surface density profiles can be transformed into $K$-band surface brightness profiles, assuming a given mass-to-light ratio (see Sect.~\ref{Sec:scaling} for more details). The mass in the remnants is basically distributed according to a bulge$+$disk structure, therefore a combined S\'{e}rsic$+$exponential function was fitted to the obtained surface density profiles. The Sersic function \citep{1968adga.book.....S} provides a good description of the bulge component in these profiles \citep[e.g.,][]{1994MNRAS.267..283A,1995MNRAS.275..874A,2001AJ....121..820G,2001A&A...368...16M,2001A&A...367..405P,2003ApJ...582..689M,2008A&A...478..353M}:

\begin{equation}
 I(r)=I_{\mathrm{eff}} \exp\,\{- b_{n}\, [ ( r/r_{\mathrm{eff}}) ^{1/n}-1] \} , \label{Eq:Sersic}
\end{equation}

\noindent where $r_{\mathrm{eff}}$ is the bulge effective radius, $I_{\mathrm{eff}}$ is the surface density at $r_{\mathrm{eff}}$, and $n$ is the S\'{e}rsic index (a parameter related to the bulge concentration). The factor $b_{n}$ is a function of $n$, which may be approximated by $b_{n}=1.9992\, n-0.3271$ in the range $1<n<10$ with an error $<0.15$\% \citep{1991A&A...249...99C,2001MNRAS.326..543G}. 

The disk is well-described by an exponential law \citep[][]{1970ApJ...160..811F}:
\begin{equation}
I(r)=I_{\mathrm{0,D}} \exp\left( -r/h_{\mathrm{D}}\right) , \label{Eq:disc}
\end{equation}

\noindent where $h_{\mathrm{D}}$ is the scale-length of the disk and $I_{\mathrm{0,D}}$ its central surface density. 

We used a least-squares fitting procedure (see more details and some examples in EM06). Residuals are below $\sim$0.2 mag along the whole radial range. The errors of the bulge and disk photometric parameters were estimated with a bootstrap method \citep{Efron93,Press94}. In Table~\ref{tab:photometricparameters}, we list the final values of the bulge and disk photometric parameters derived from these fits for all remnants, assuming the scalings and the mass-to-light ratio commented in Sect.~\ref{Sec:scaling}. We remark that these models are scalable in size, mass, and velocity, i.e., these parameters can be scaled to a different galaxy simply by assuming different mass and size units in the simulations.

\section{Scaling of the models}
\label{Sec:scaling}

These models are scalable in mass, size, and velocity. The primary model can be scaled to an S0b galaxy comparable to the Milky Way (MW) by considering $R=4.5$\,kpc, $v=220$\,km s$^{-1}$, and $M=5.1\times 10^{10} M_\odot$ as the units of length, velocity, and mass, respectively (the corresponding time unit would then be 20.5\,Myr). The primary S0c galaxy can be scaled to NGC\,253 considering $R=6.8$\,kpc, $v=510$\,km s$^{-1}$, and $M=2.6\times 10^{11} M_\odot$ as units (the time unit is 11.7\,Myr). Assuming these scalings, the time period computed after the full merger ranges from 0.4 to 1.1 Gyr depending on the model. 

To compare our models with observational data, we need to transform the mass in the models to luminosity. This comparison is only realistic for data that basically trace the stellar mass in galaxies, which these models do, because they do not include gas and star formation effects. Therefore, we adopted the NIRS0S survey as the observational reference in this study \citep[NIRS0S for Near-IR S0 galaxy Survey, see][]{2011MNRAS.418.1452L}. This sample contains photometric data in the $K$ band for $\sim 180$ early-type nearby galaxy disks ($\sim 120$ are S0s), and multi-component decompositions of them have been published by L10. The advantage of these near-IR decompositions is that they are only weakly affected by dust and star formation, which makes them excellent tracers of  stellar mass. 
 
We assumed a mass-to-light ratio in the $K$ band typical of early-type galaxies: $M_\mathrm{stars}/L_\mathrm{K}\sim 1$\, $M_{\odot}/L_{\odot,\mathrm{K}}$ \citep{2000Ap.....43..145R,2004MNRAS.347..691P}. Therefore, a direct equivalence between the surface density and the $K$-band surface brightness can be established in our models, for a given mass scaling (for more details, see Paper I). 

We assumed the scalings provided at the beginning of this section throughout the paper to facilitate the comparison with real data, but we stress that they can be modified. The primary S0b galaxy is scaled to the mass and size of the MW, while galaxy S0c is scaled to NGC\,253. The real young stellar populations and dust content of these two galaxies (which are spirals) are not expected to noticeably affect their stellar masses and scale-lengths (see Paper I). 

For the MW, a radius $R=13.9$\,kpc \citep{2011ApJ...733L..43M} and a total magnitude $K_\mathrm{T}=-23.8$\,mag were adopted. This last value has been estimated through the conversion of the MW $B$-band total magnitude  \citep[$B_\mathrm{T}=-20.3$\,mag, see][]{1997A&A...326..897H} to the $K$ band, assuming the typical colour of nearby early-type galaxies in the 2MASS Large Galaxy Atlas \citep[$B-K \sim 4$\,mag, ][]{2003AJ....125..525J}. With these values, the central surface brightness of the disk of our S0b primary galaxy is $\mu_\mathrm{0,D} = 17.10$\,mag/arcsec$^2$ in the $K$ band (see Table\,\ref{tab:photometricparameters}), a value very similar to the one reported by \citet[]{2002ApJ...581.1013B} for the MW ($\mu_\mathrm{0,D}(K) = 17$\,mag/arcsec$^2$). 

For NGC\,253, we considered a radius $R=14.2$\,kpc and a total magnitude $K_\mathrm{T}=-23.8$\,mag, also in the $K$ band. These physical quantities are derived from the data provided by \citet[]{2003AJ....125..525J}, assuming the apparent $K$-band magnitude and the distance to this galaxy reported by these authors ($m_K=3.77$\,mag at $D=3.5$\,Mpc). With these values, the effective surface brightness of the bulge of our S0c primary galaxy is $\mu_\mathrm{eff} = 15.97$\,mag/arcsec$^2$ in the $K$ band, which coincides with the value reported by \citet{2003AJ....125..525J} for NGC\,253 (see Table\,\ref{tab:photometricparameters}).

\section{Results}
\label{Sec:results}

In this section, we present the evolution induced by the simulated mergers in the photometric planes studied by L10. The photometric relations between the global structural bulge-disk parameters ($n$, $r_\mathrm{eff}$, $h_\mathrm{D}$, $r_\mathrm{eff}/h_\mathrm{D}$, and $B/T$) have been studied in Paper I, therefore we restrict ourselves here to the photometric relations involving other parameters of the bulge (Sect.~\ref{Sec:bulgeall}), of the disk (Sect.~\ref{Sec:disk}), and one parameter of the bulge with one of the disk (Sect.~\ref{Sec:bulgedisk}) that were not presented previously. We also compare our results with the evolution triggered by the merger experiments of high-density dSph satellites (similar to compact ellipticals, cE) onto an S0b performed by \citet{2001A&A...367..428A}. 

We remark that the models can be moved to different locations in most photometric planes by simply assuming different mass and size scaling. As an example, we have marked the region that could be achievable by our models if one were to use a different scaling only in two photometric planes (those that compare the $K$-band magnitudes of different components in the galaxies). This is because only mass scaling affects them, which makes the trend of the scaling quite distinguishable from the evolutionary trend induced by the mergers. However, in the photometric planes where mass and size scalings can be combined, the scaling trend is not as intuitive, therefore we prefer to obviate them for clarity.  We adopted the model scaling reported in Sect.~\ref{Sec:scaling} in all figures to facilitate the comparison between the models and the data.

\begin{figure}[t!]
\center
\includegraphics[width = 0.5\textwidth]{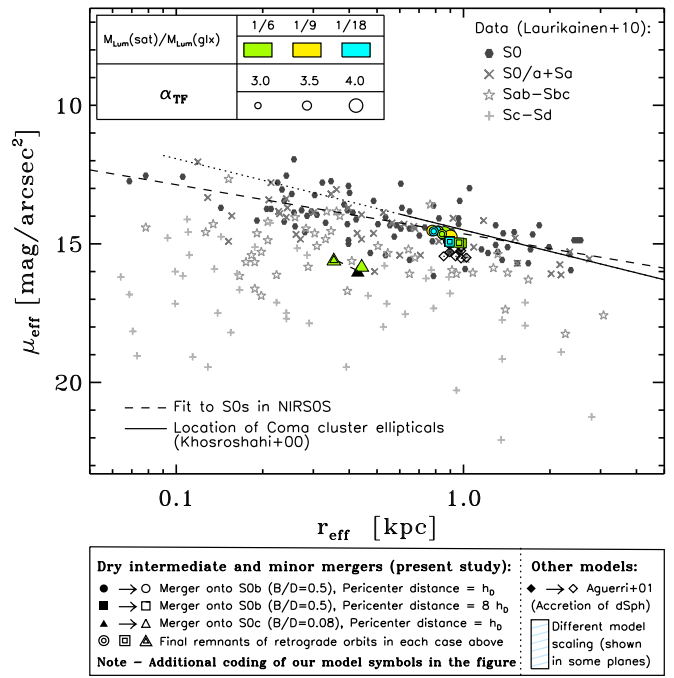} 
\caption{Evolution in the Kormendy relation ($\mu_\mathrm{eff}$ vs.~$r_\mathrm{eff}$ in the bulges) driven by our merger experiments, compared to the observational distributions of S0s and spirals by L10. Growth vectors have been plotted, starting at the location of the original galaxy model and ending at the $\mu_\mathrm{eff}$-$r_\mathrm{eff}$ values of the remnants. The growth vectors corresponding to the merger models with dense spheroidal satellites by \citet{2001A&A...367..428A} are represented. In both cases, the evolution induced by the mergers in the plane is negligible. The scalings as we detailed in Sect.~\ref{Sec:scaling} have been assumed for all plots, but note that the models can be scaled in mass and size. The location of Coma cluster ellipticals from \citet{2000ApJ...531L.103K} is marked with a solid line. Consult the legend for the observational data and the models in the figure.  }\label{fig:kormendy}
\end{figure}

\subsection{Photometric relations of the bulge parameters}
\label{Sec:bulgeall}

\subsubsection{Kormendy relation}
\label{Sec:kormendy}

There is a strong correlation between the surface central brightness of early-type galaxies and their effective radius. This is 
the so-called Kormendy relation, initially found for elliptical galaxies \citep{1977ApJ...218..333K}. Nevertheless, several authors have shown that bulges of S0 galaxies also follow the same relation \citep[see e.g.][]{2010A&A...521A..71M}, although there is a small offset of the relation between S0 bulges fainter than $M_{k}=-22.7$\,mag and ellipticals. In this sense, the bulges of late-type galaxies show a larger offset with respect to the Kormendy relation than ellipticals \citep[see][]{1979ApJ...234..435B,1985ApJS...59..115K,1994MNRAS.267..283A,2004AJ....127.1344A,2006A&A...446..827R}.

In Fig.\,\ref{fig:kormendy}, we show the evolution induced by our merger experiments in the Kormendy relation. Our minor mergers move the bulges of the initial galaxies (which lie slightly below the region occupied by S0s because of the selected scalings, see Sect.~\ref{Sec:scaling}) towards the location of the bulges of real S0 galaxies. This evolution can be explained by considering the increment in the bulge mass (i.e., in its luminosity) induced by the mergers (see Paper I), because \citet{2008A&A...491..731N} demonstrated that the distribution of real galaxies in the $\mu_\mathrm{eff}$ - $\log(r_\mathrm{eff}$) plane basically depends on the galaxy luminosity (and thus, on their  masses). In general, the dispersion induced by the mergers in this photometric plane is compatible with the dispersion observed in the bulges of real S0s reported by L10.

\begin{figure}[t!]
\center
\includegraphics[width = 0.5\textwidth,  bb= 10 98 481 481, clip]{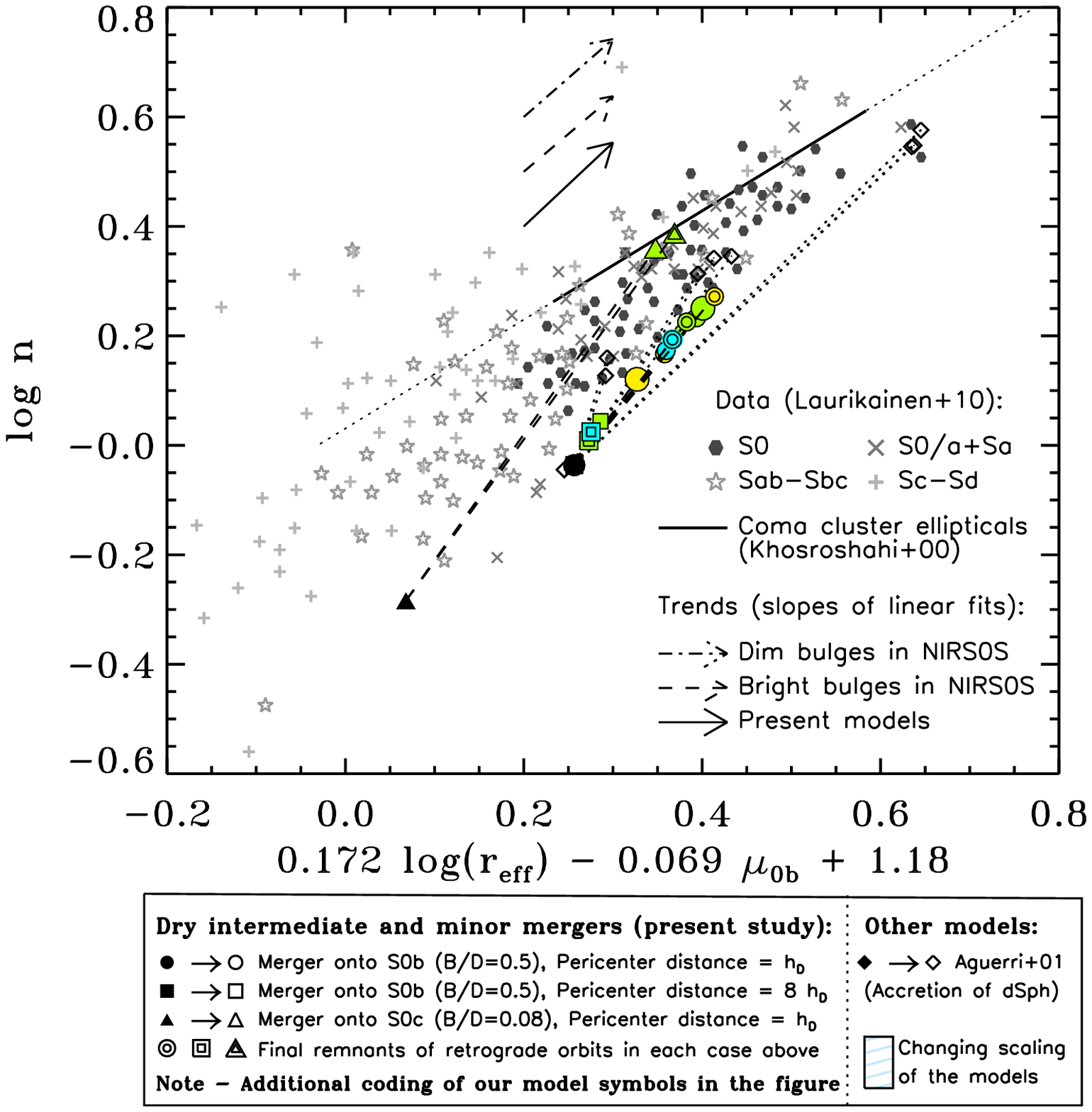} 
\caption{Growth vectors in the photometric plane of the bulges (which relates the three bulge parameters: $n$, $\mu_\mathrm{eff}$, and $r_\mathrm{eff}$)  driven by our merger experiments, compared to the observational distributions of S0s and spirals by L10. The dashed and dotted-dashed arrows indicate the trends observed by L10 for bright ($M_K (\mathrm{bulge}) < -22.7$\,mag) and faint bulges ($M_K (\mathrm{bulge}) > -22.7$\,mag) of the S0s in the NIRS0S survey. The solid arrow marks the trend found in our models. Consult the legend for the observational data in the figure. The legend for the models is the same as in Fig.\,\ref{fig:kormendy}. 
}\label{fig:fp}
\end{figure}

\subsubsection{Photometric plane}
\label{Sec:photometricplane}

Bulges of S0 galaxies and ellipticals are distributed in a thin plane called the fundamental plane \citep[FP,][]{1987ApJ...313...59D,1987ApJ...313...42D,2002MNRAS.335..741F}. The FP relates the structure and the kinematics of the galaxies, and it is related with the virial theorem. It has been shown that the Sersic $n$ parameter strongly correlates with the central velocity dispersion of elliptical galaxies \citep[see e.g.][]{2001AJ....122.1707G}. This correlation between $n$ and $\sigma$ provides another representation of the FP, called the photometric plane (PHP).

The PHP of early-type galaxies has been extensively investigated in the literature \citep{1999MNRAS.309..481L,2000A&A...353..873M,2002MNRAS.334..859G,2000ApJ...531L.103K,2004MNRAS.349..527K,2012MNRAS.420.2835K}. In Fig.\,\ref{fig:fp} we show the PHP for bulges of S0 and spirals from L10. Note that bulges of S0 and spirals define the same PHP. The slopes of the linear fits of the PHP of bright ($M_{K}(\mathrm{bulge})<-22.7$\,mag) and faint ($M_{K}(\mathrm{bulge})>-22.7$\,mag) bulges are similar. Nevertheless, S0 bulges are located in the region with higher $\log(n)$ values. We have overplotted the results from our merger models. The mass growth induced by the mergers in the bulges is the reason why the final remnants have higher $n$ values than originally, placing them in the region of the PHP occupied by S0 bulges. In addition, the slope of the linear fit of the PHP of our simulated bulges is similar to the slopes of faint and bright real S0 bulges. This indicates that minor mergers move the S0 bulges along the observed PHP of S0s.

\begin{figure*}[t!]
\center
\includegraphics[width = 0.49\textwidth,  bb= 10 98 481 481, clip]{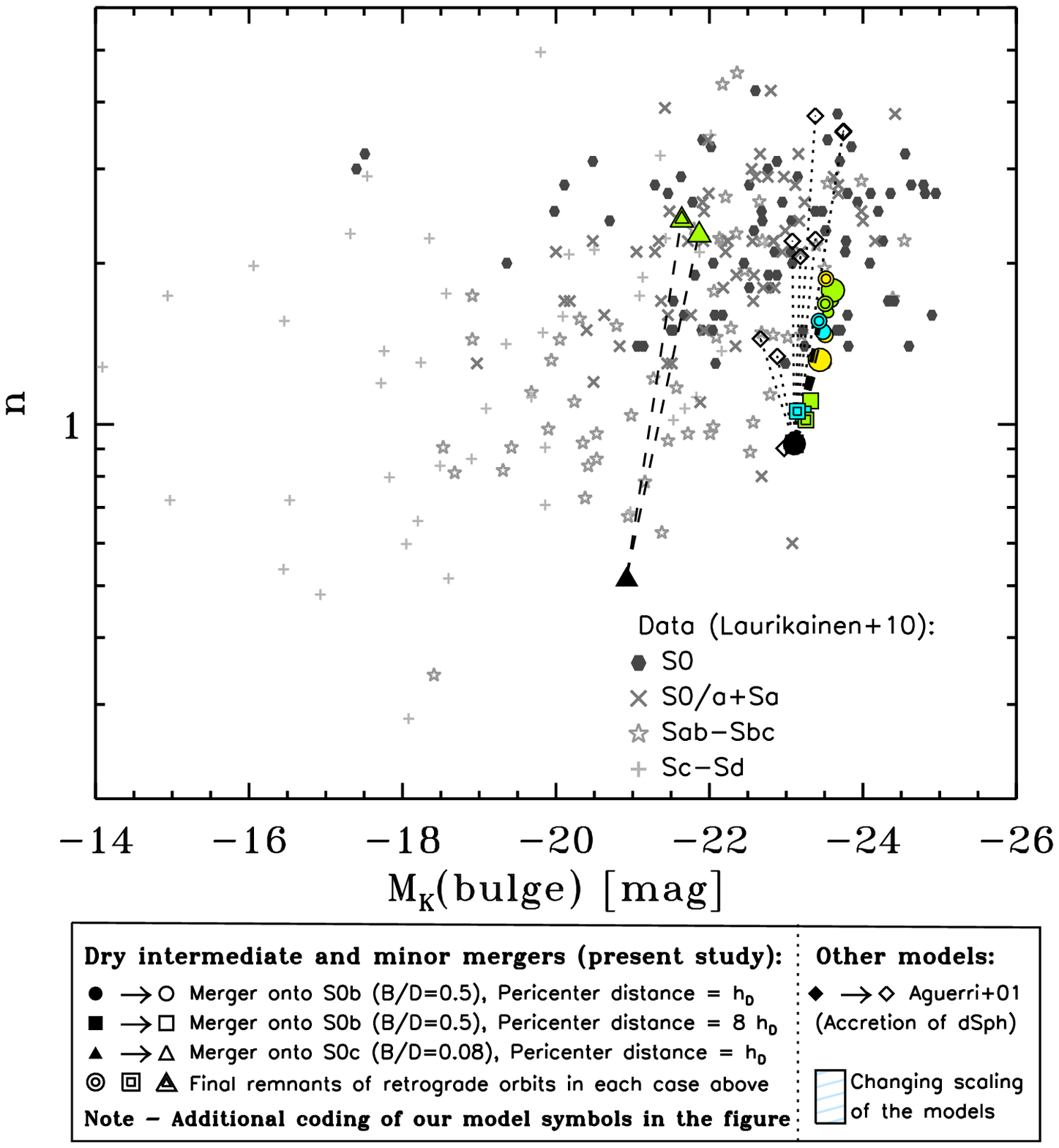} 
\includegraphics[width = 0.49\textwidth,  bb= 10 98 481 481, clip]{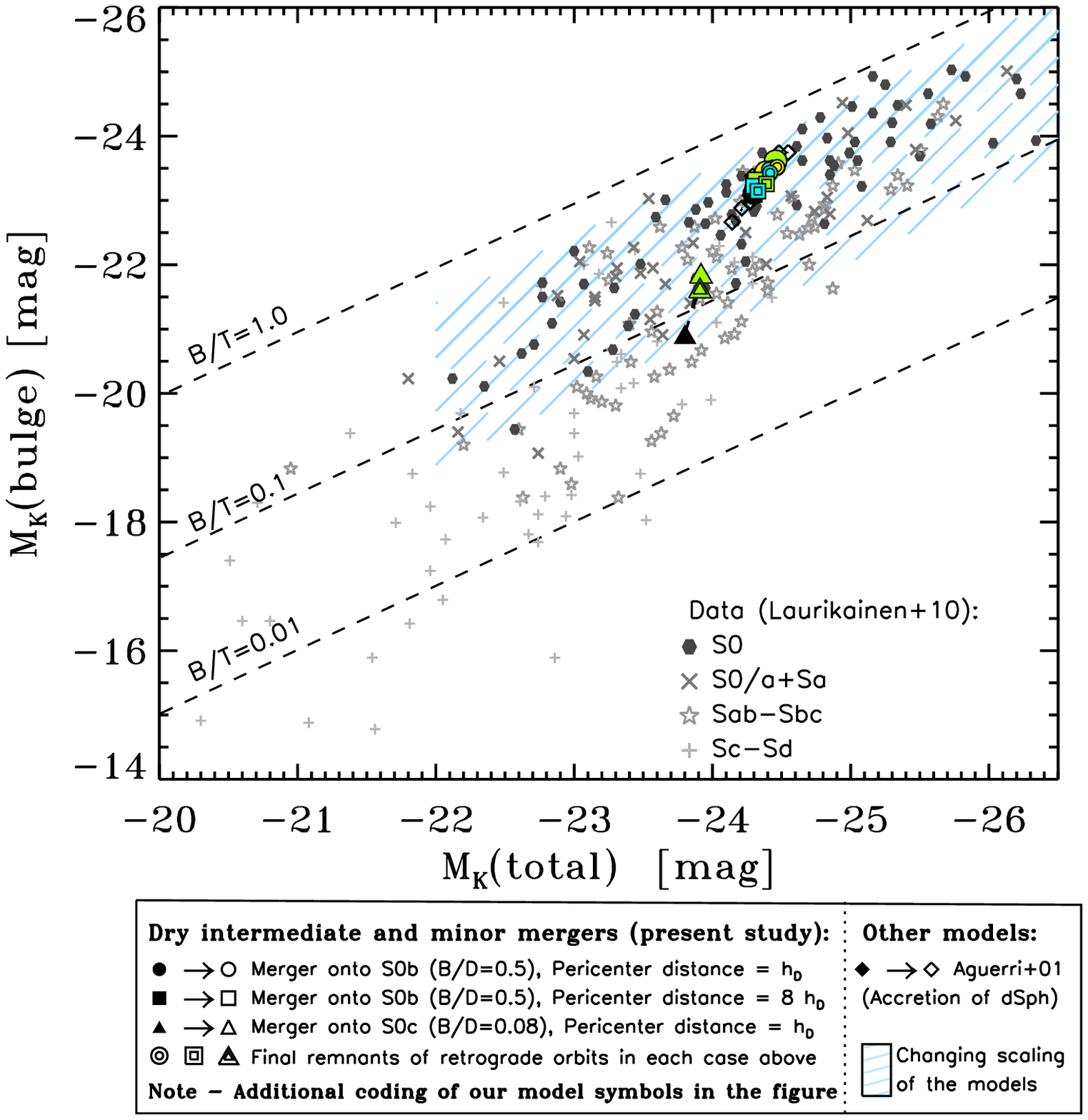} 
\caption{Growth vectors in the $n$-$M_K(\mathrm{bulge})$ (\emph{left}) and $M_K(\mathrm{bulge})$-$M_K(\mathrm{total})$  (\emph{right}) planes driven by our minor merger experiments, compared to observational data. The dashed lines denote $B/T = 0.01$, 0.1, and 1.0 from L10. The blue shaded region \emph{in the right panel} indicates the locations in the plane that are covered by our models simply by using a different mass scaling. Consult the legend for the observational data in the figure. The legend for the models is the same as in Fig.\,\ref{fig:kormendy}. 
}\label{fig:nmkbulge}
\end{figure*}

\begin{figure*}[t!]
\center
\includegraphics[width = 0.49\textwidth,  bb= 10 98 481 481, clip]{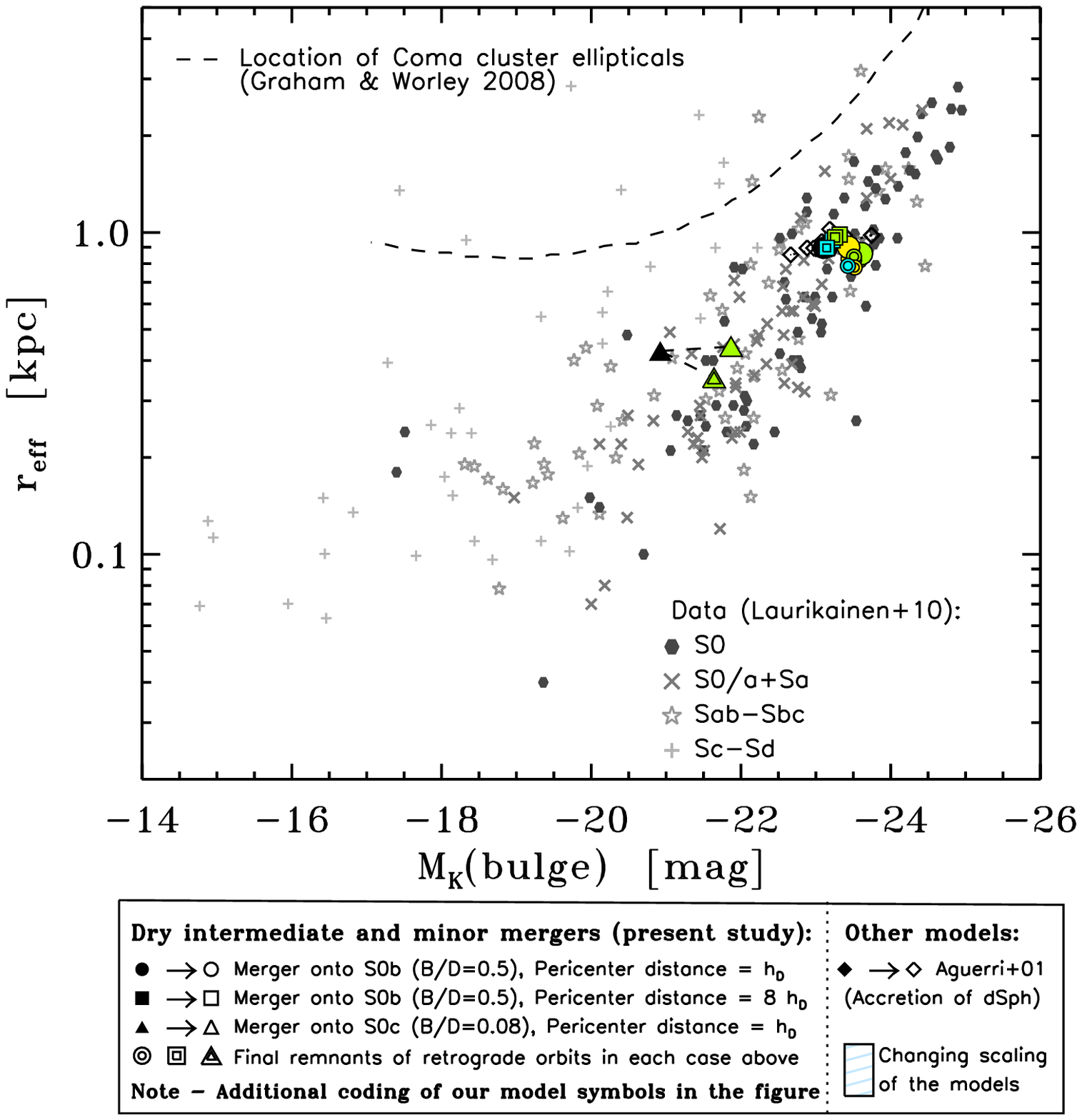} 
\includegraphics[width = 0.49\textwidth,  bb= 10 98 481 481, clip]{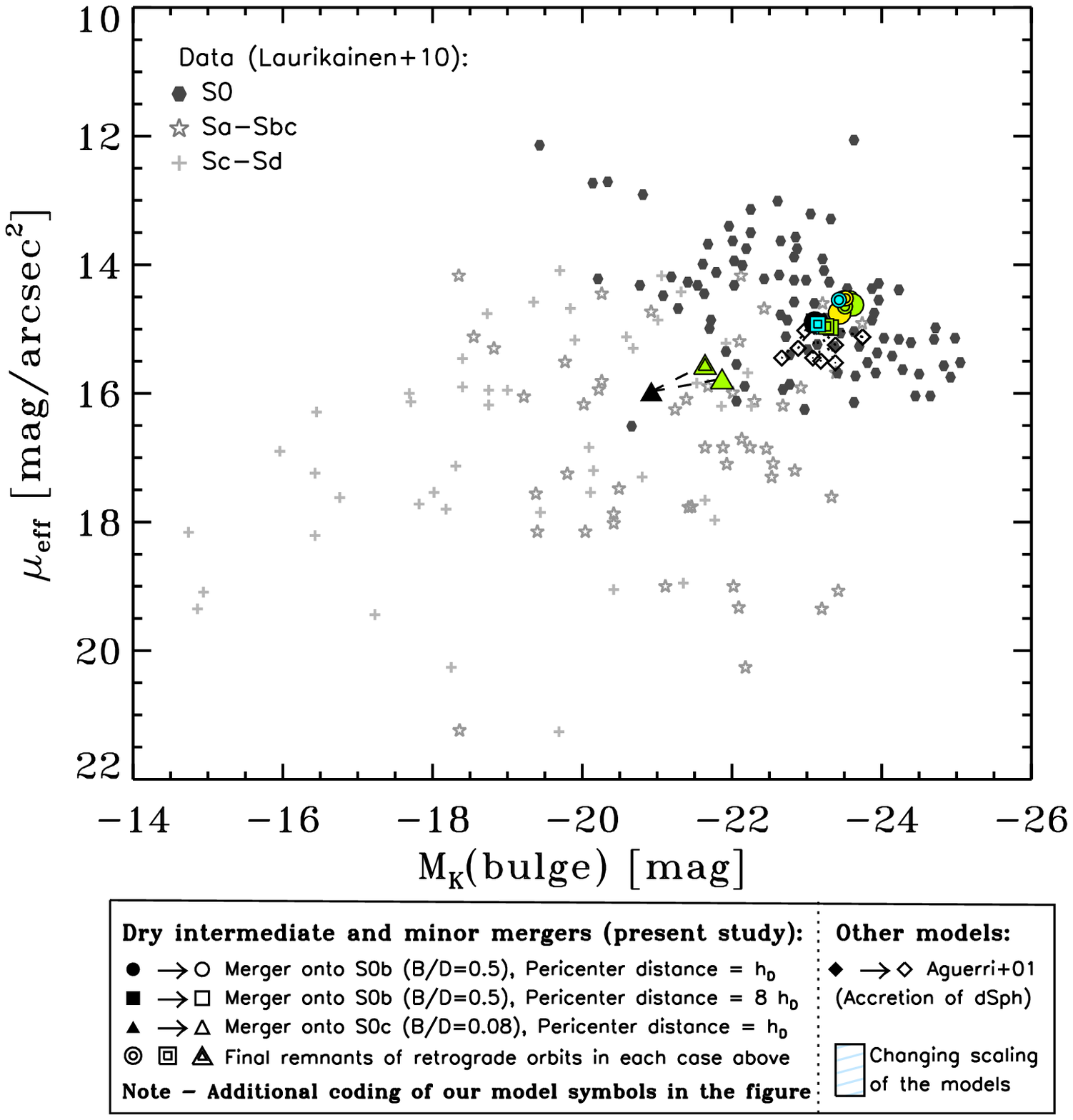} 
\caption{Evolution in the $r_\mathrm{eff}$-$M_K(\mathrm{bulge})$  (\emph{left})  and $\mu_\mathrm{eff}$-$M_K(\mathrm{bulge})$  (\emph{right})  planes driven by our minor merger experiments, compared to real data. The dashed line \emph{in the left panel} marks the location of Coma cluster ellipticals by \citet{2008MNRAS.388.1708G}. Consult the legend for the observational data in the figure. Growth vectors have been plotted for the models, but the evolution induced by the mergers in these planes is negligible. The legend for the models is the same as in Fig.\,\ref{fig:kormendy}. 
}\label{fig:bulge}
\end{figure*}

\begin{figure*}[t!]
\center
\includegraphics[width = 0.49\textwidth,  bb= 10 98 481 481, clip]{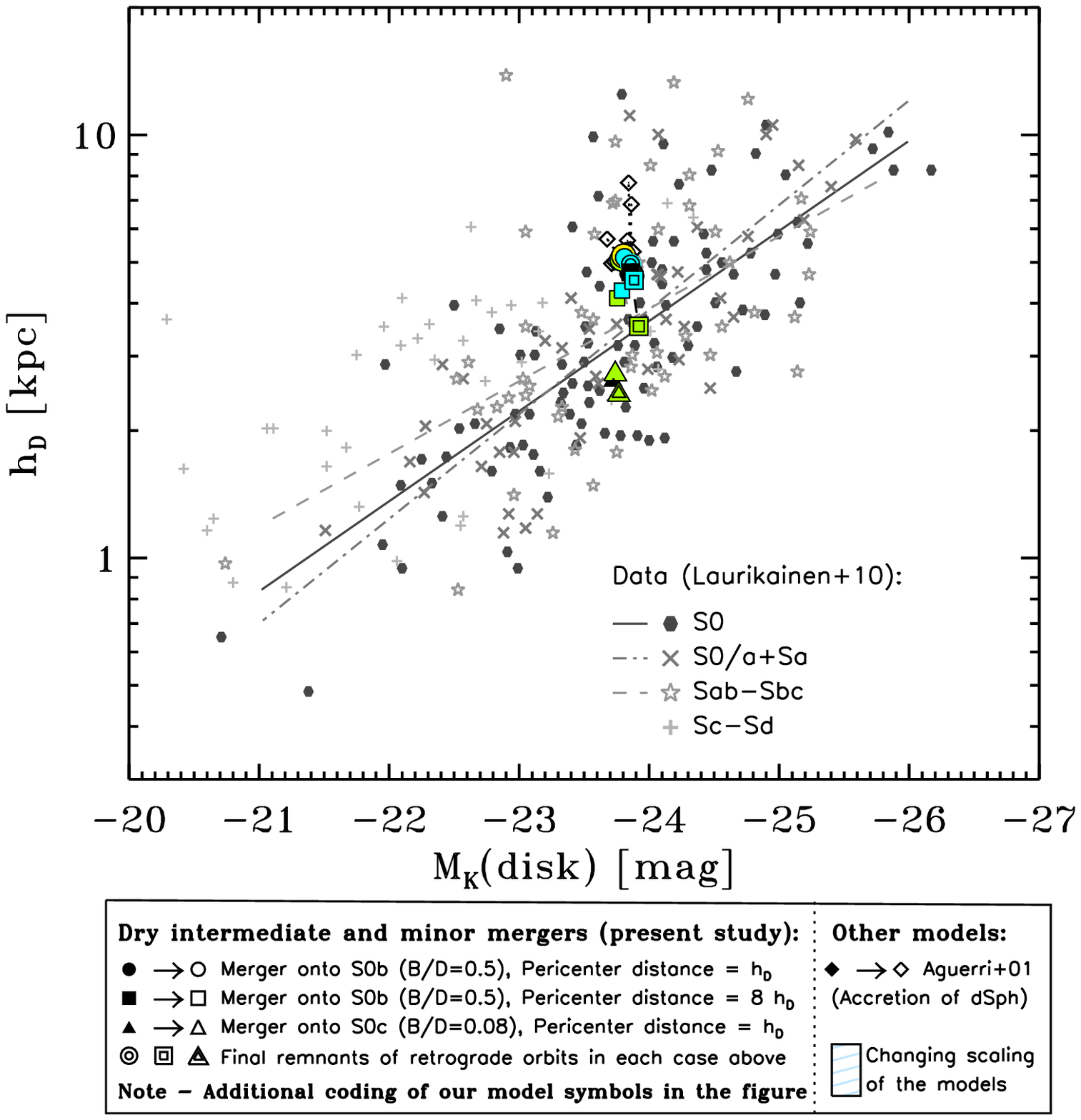} 
\includegraphics[width = 0.49\textwidth,  bb= 10 98 481 481, clip]{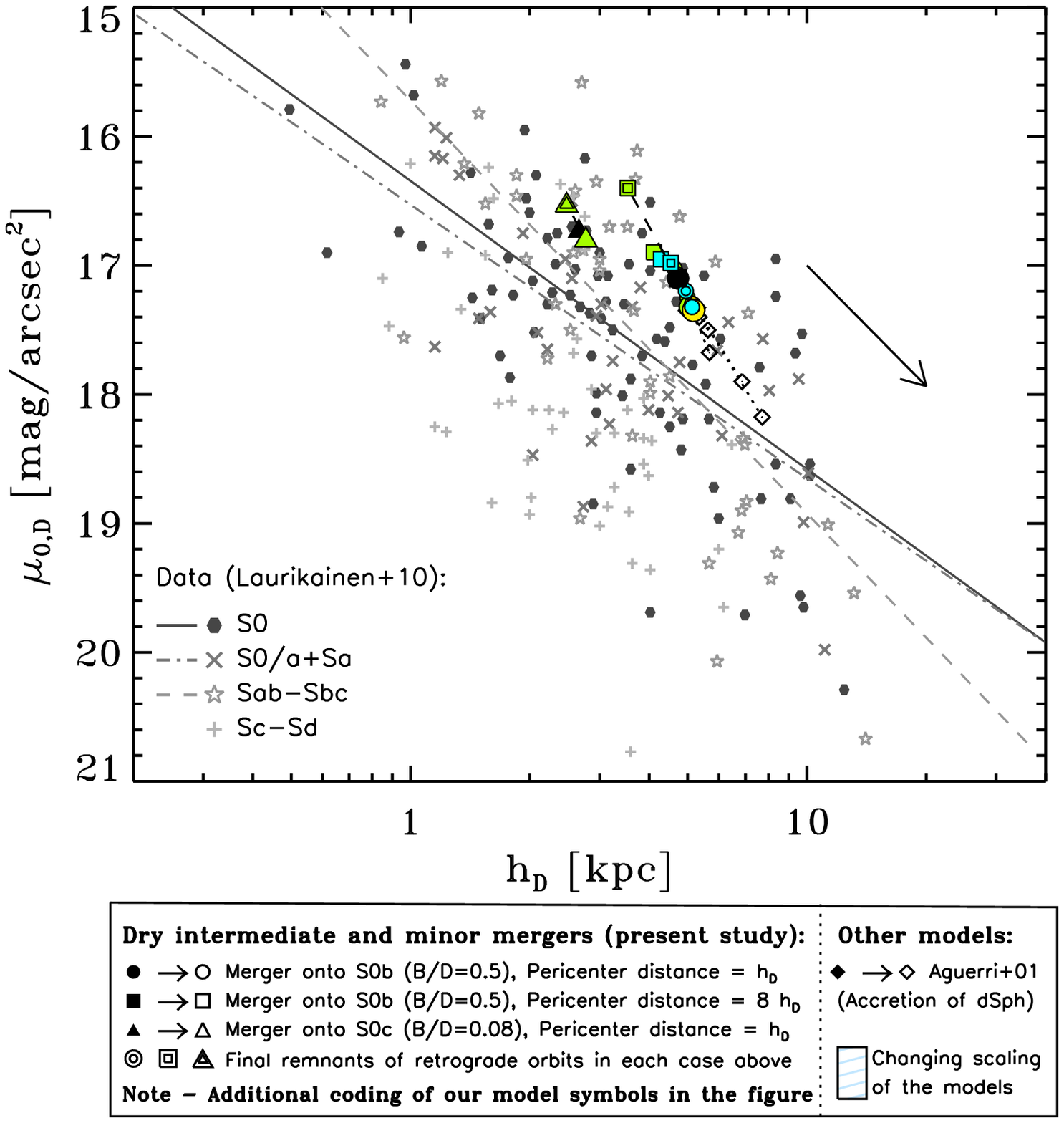} 
\caption{Growth vectors in the $h_\mathrm{D}$-$M_K(\mathrm{disk})$  (\emph{left}) and $\mu_\mathrm{0,D}$-$h_\mathrm{D}$ planes (\emph{right})  driven by our minor merger experiments, compared to real data. The straight lines represent the best fits for S0s (solid lines), S0/a$+$Sa (dotted-dashed lines), and Sab-Sbc galaxies (dashed lines) obtained by L10, which all have statistically significant rank correlation coefficients. The arrow \emph{in the right panel} indicates the slope of the trend drawn by the merger models in the plane. Consult the legend for the observational data in the figure. The legend for the models is the same as in Fig.\,\ref{fig:kormendy}. 
}\label{fig:disk}
\end{figure*}

\subsubsection{Photometric relations involving the bulge magnitude}
\label{Sec:bulge}

Figure~\ref{fig:nmkbulge} shows the relation of the absolute bulge magnitude ($M_K(\mathrm{bulge})$) with the Sersic index of the bulge ($n$) and with the total magnitude of the galaxy ($M_K(\mathrm{total})$) in all the remnants, compared with observational data from L10. The intermediate and minor mergers explored so far are able to increase the bulge magnitude by up to $\sim 1$\,mag in most cases (assuming the scalings presented in Sect.~\ref{Sec:scaling}, see also Table\,\ref{tab:photometricparameters}). As stated in EM06 and EM11, this is due not only to the aggregation of material from the satellite to the central parts of the remnant, but also to the effect of the inward motion of disk material (see Sect.~\ref{Sec:discussion}). The increase in $n$ is also general in all cases. The amount of change depends on the orbital parameters and, less evidently, on the initial density contrast (see Paper I for details).

We can express the bulge total luminosity in terms of the Sersic parameters as $L = k_{\rm L} I_{\rm eff}r^{2}_{\rm eff}$, with $I_{\rm eff}$ the flux at $r_\mathrm{eff}$, and $k_{\rm L}$ a function of the Sersic index $n$. Thus a correlation between the Sersic index $n$ and $M_K(\mathrm{bulge})$ should be expected. Although a number of authors do find such correlation for disk galaxies \citep[see, e.g.,][]{2001MNRAS.326..543G}, L10 claim that no such correlation is found for S0s. The left panel of Fig.~\ref{fig:nmkbulge} shows our results in the $n$-$M_K(\mathrm{bulge})$ plane over-imposed on the observational data presented by L10. Our minor mergers are effective at increasing the value of $n$ but less so at changing the $K$-band bulge absolute magnitude. This moves the remnants towards higher $n$ values and (slightly) higher bulge magnitudes, which increases the dispersion of the remnants in the plane. 

In other words, according to this mechanism, if we start from a well-correlated set of progenitors, we may end with a less correlated distribution of remnants in the plane, accumulated at the region where most S0s reside in the diagram. Therefore, our experiments indicate that dry intermediate and minor mergers agree with the observational fact that S0s do not present a clear $n$-$M_K(\mathrm{bulge})$ correlation and also with their observational distribution. 

The right panel of Fig.~\ref{fig:nmkbulge} shows the $M_{K}(\mathrm{bulge})$ vs. $M_{K}(\mathrm{total})$ relation for our merger simulations after scaling compared to observational data. In all cases, we see that both the total and bulge $K$-band absolute magnitudes increase after the merger event, reflecting the fact that the total mass and the bulge mass of the system have increased. The brightening is higher in the bulge magnitude than in the total magnitude, indicating that the relative mass increase is not the same in the two components. Indeed, mergers tend to increase the bulge mass, but slightly decrease the disk mass (see Sect.\,\ref{Sec:bulgedisk} and Fig.\,\ref{fig:bulgedisk}). This means that the minor mergers increase the final $B/T$ ratio, as already noted in Paper I.

S0s show a well-defined relation in the $M_{K}(\mathrm{bulge})$-$M_{K}(\mathrm{total})$ plane that disperses and even vanishes when other later Hubble types are considered (compare the data symbols in the right panel of Fig.~\ref{fig:nmkbulge}).  L10 found that the bulges of bright spirals tend to behave similarly to those of S0s. In this sense, it is interesting to note that dry intermediate and minor mergers would give rise to brighter bulges and contribute to accumulating the remnants in the region populated by S0s in the diagram and building up this relation, in agreement with observations.

Figure~\ref{fig:bulge} shows the other two correlations that might be expected from the Sersic fitting of the bulge parameters: $r_{\rm eff}$ vs. $M_{K}(\mathrm{bulge})$ and $\mu_{\rm eff}$ vs. $M_{K}(\mathrm{bulge})$. While minor mergers tend to brighten the $K$-band absolute magnitude of the bulge, it is not so clear that this increment is translated into an expansion of the effective radius (see the left panel of the figure). The change induced in the bulge effective radius by these mergers is negligible. Therefore, the final bulge is more luminous than the progenitor bulge in all cases but has a similar scale-length, which agrees with the observed accumulation of S0s on a strip at the right side of this diagram. As a secondary effect, the remnants tend to have brighter effective surface brightness values than the progenitors, and thus a sequence of these minor mergers would move the bulges towards the top right corner of the right panel of Fig.~\ref{fig:bulge}, where S0s accumulate as well. Therefore, the evolution induced by these mergers is completely compatible with the observed distribution of S0s in these photometric planes. Assuming a different scaling for our initial S0 models, the final remnants can reproduce the region populated by S0s in the $r_{\rm eff}$-$M_{K}(\mathrm{bulge})$ diagram (left panel of the figure), but they could hardly promote a galaxy initially at the S0s region to the location of the Coma cluster ellipticals in this photometric plane.

The change in $r_\mathrm{eff}$ and $\mu_\mathrm{eff}$ induced by the merger differs depending on the pericentre distance: 

\begin{enumerate}
 \item Mergers with short-pericentre orbits tend to decrease $r_\mathrm{eff}$, brightening $\mu_\mathrm{eff}$ (i.e., they tend to compress the bulges).
 \item Mergers with long-pericentre orbits do the opposite: they rise $r_\mathrm{eff}$ and dim $\mu_\mathrm{eff}$ (i.e., expand the bulges).
\end{enumerate}

Therefore, long-pericentre orbits tend to give rise to more expanded bulges than initially, whereas short-pericentre orbits tend to concentrate them. We will show that analogue, but opposite, trends with the pericentre distance are found for the disk parameters $h_\mathrm{D}$ and $\mu_\mathrm{0,D}$ (see Sect.~\ref{Sec:disk}). 

In general, we do not find clear trends in the photometric scaling relations involving bulge parameters with the mass ratio of the encounter, the central satellite density, or the spin-orbit coupling (see Figs.~\ref{fig:kormendy}-\ref{fig:bulge}). In the experiments with an original primary S0b galaxy, the encounters with long-pericentre orbits induce weaker changes to the bulge photometric parameters than those with short pericentres. Additionally, the evolution triggered in the models with a primary S0c galaxy is more noticeable than in those with an S0b for the same initial conditions, as already noted in paper I. 

Summarizing, the bulges of our remnants are more similar to the bulges of spirals than to ellipticals regarding the $M_K(\mathrm{bulge})$-$r_\mathrm{eff}$ relation, the photometric plane, and the Kormendy relation, in agreement with the results reported by L10 for real S0s. We can therefore conclude from this section that dry intermediate and minor mergers onto S0s make these galaxies  evolve along the photometric scaling relations of S0 bulges.

\subsection{Photometric relations of the disk parameters}
\label{Sec:disk}

\citet{2010MNRAS.405.1089L} confirmed the strong correlations between the disk scale-length ($h_\mathrm{D}$) and other disk parameters previously reported by other authors for S0s \citep{2008A&A...487..555M,2008A&A...478..353M}. We represent the evolution driven by our merger experiments in these two photometric planes in Fig.~\ref{fig:disk} and compare them to the trends observed by L10. Although the evolution induced by the mergers agrees completely with the observational distribution of S0s in the two diagrams, the trends induced by mergers differ. 

Observationally, $h_\mathrm{D}$ increases as $M_K(\mathrm{disk})$ brightens in S0s (see the left panel of Fig.~\ref{fig:disk}). The figure shows that the evolution induced by the mergers is coherent with this trend, not because mergers move galaxies to build up this trend, but because they only trigger a negligible change in both $h_\mathrm{D}$ and $M_K(\mathrm{disk})$. If the $h_\mathrm{D}$-$M_K(\mathrm{disk})$ relation were established in galaxies at early times, dry intermediate and minor mergers would preserve it and would merely cause a slight dispersion. The available data support this scenario, because the relation between the disk magnitude and its scale-length is similar in all galaxy types from Sc to S0. Indeed, the distributions of the different types overlap in the diagram. 

In contrast, the evolution induced by the mergers in the $\mu_\mathrm{0,D}$-$h_\mathrm{D}$ plane agrees with the observational distribution of S0s not because they affect negligibly to these parameters, but because they evolve galaxies to build up the observed trend (see the right panel of Fig.~\ref{fig:disk}). Observationally, $\mu_\mathrm{0,D}$ is observed to decrease with rising $h_\mathrm{D}$ in all types (and in particular, in S0s). Our models consistently reproduce the observed trend. Although all observed morphological types follow a well-defined trend and overlap in this plane (similarly to the one shown in the left panel), the mergers now drive a well-defined evolutionary trend in the $\mu_\mathrm{0,D}$-$h_\mathrm{D}$ plane, instead of just introducing a slight dispersion (as happens in the $h_\mathrm{D}$-$M_K(\mathrm{disk})$ relation, left panel of the figure). 

\begin{figure}[t!]
\center
\includegraphics[width = 0.5\textwidth,  bb= 10 98 481 481, clip]{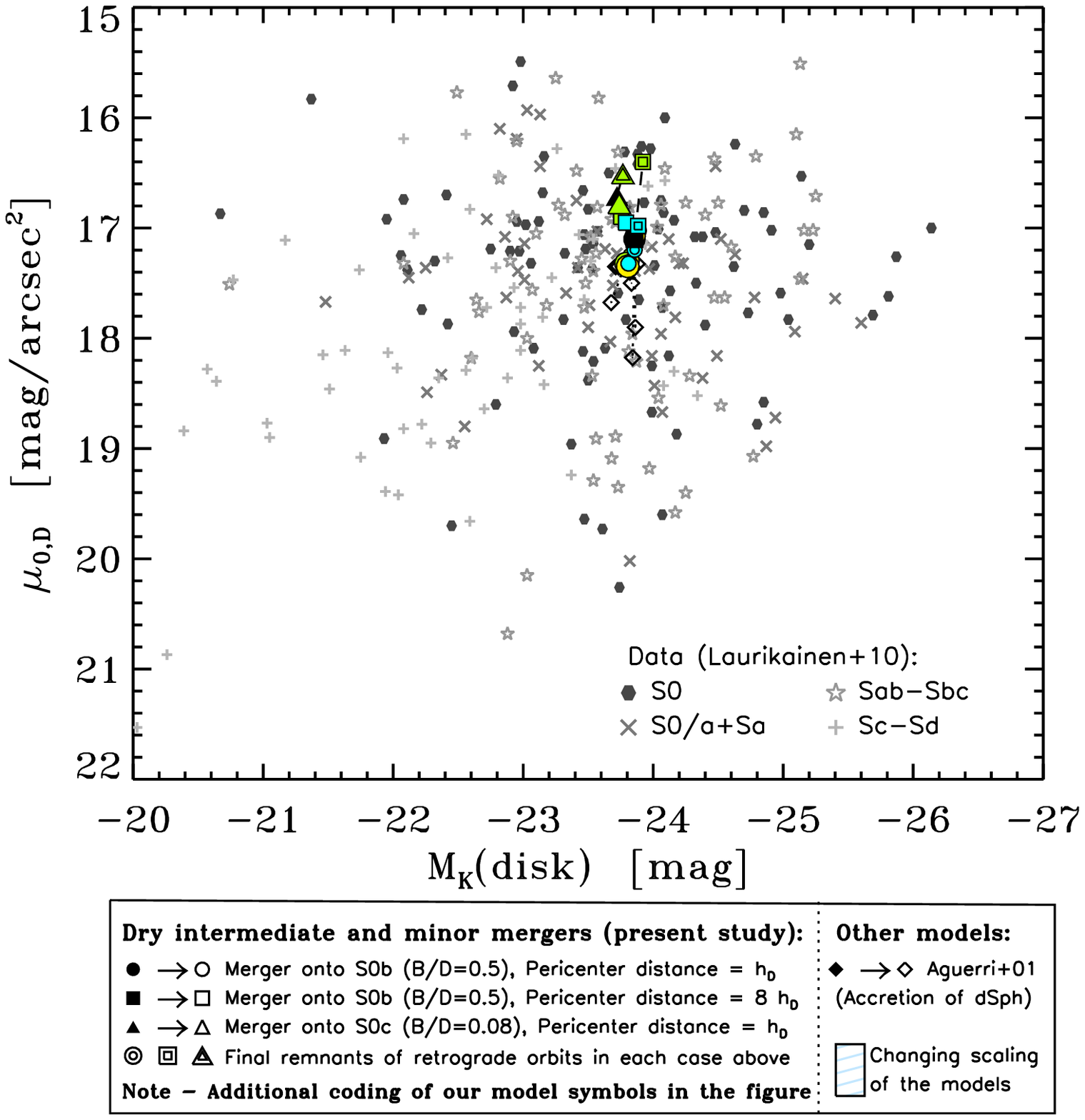} 
\caption{Evolution in the $\mu_\mathrm{0,D}$-$M_K(\mathrm{disk})$ plane driven by our minor merger experiments, compared to the observational distributions of S0s and spirals by L10. Consult the legend for the observational data in the figure. The legend for the models is the same as in Fig.\,\ref{fig:kormendy}. 
}\label{fig:disk2}
\end{figure}

\begin{figure*}[t!]
\center
\includegraphics[width = 0.49\textwidth,  bb= 10 98 481 481, clip]{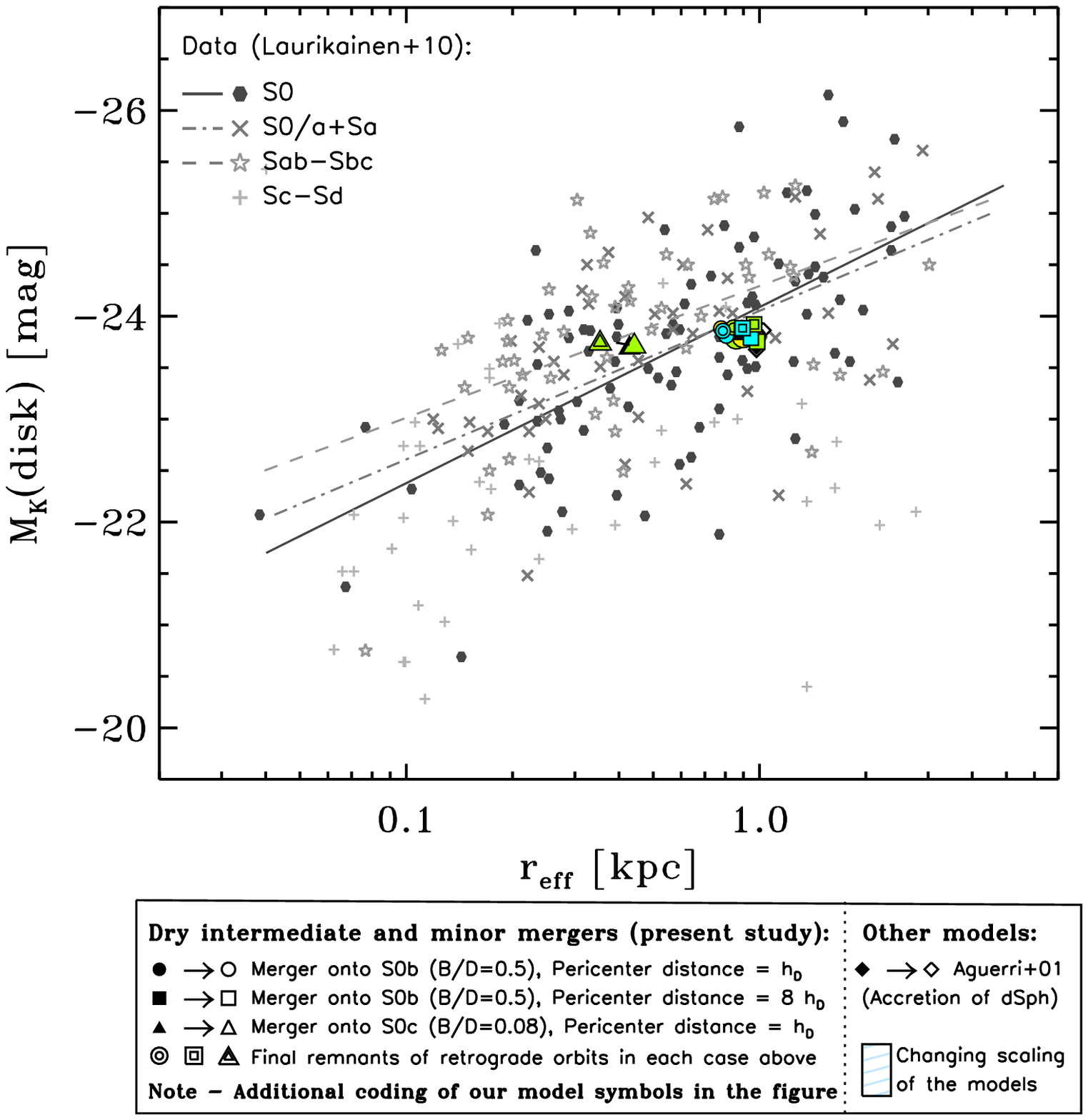} 
\includegraphics[width = 0.49\textwidth,  bb= 10 98 481 481, clip]{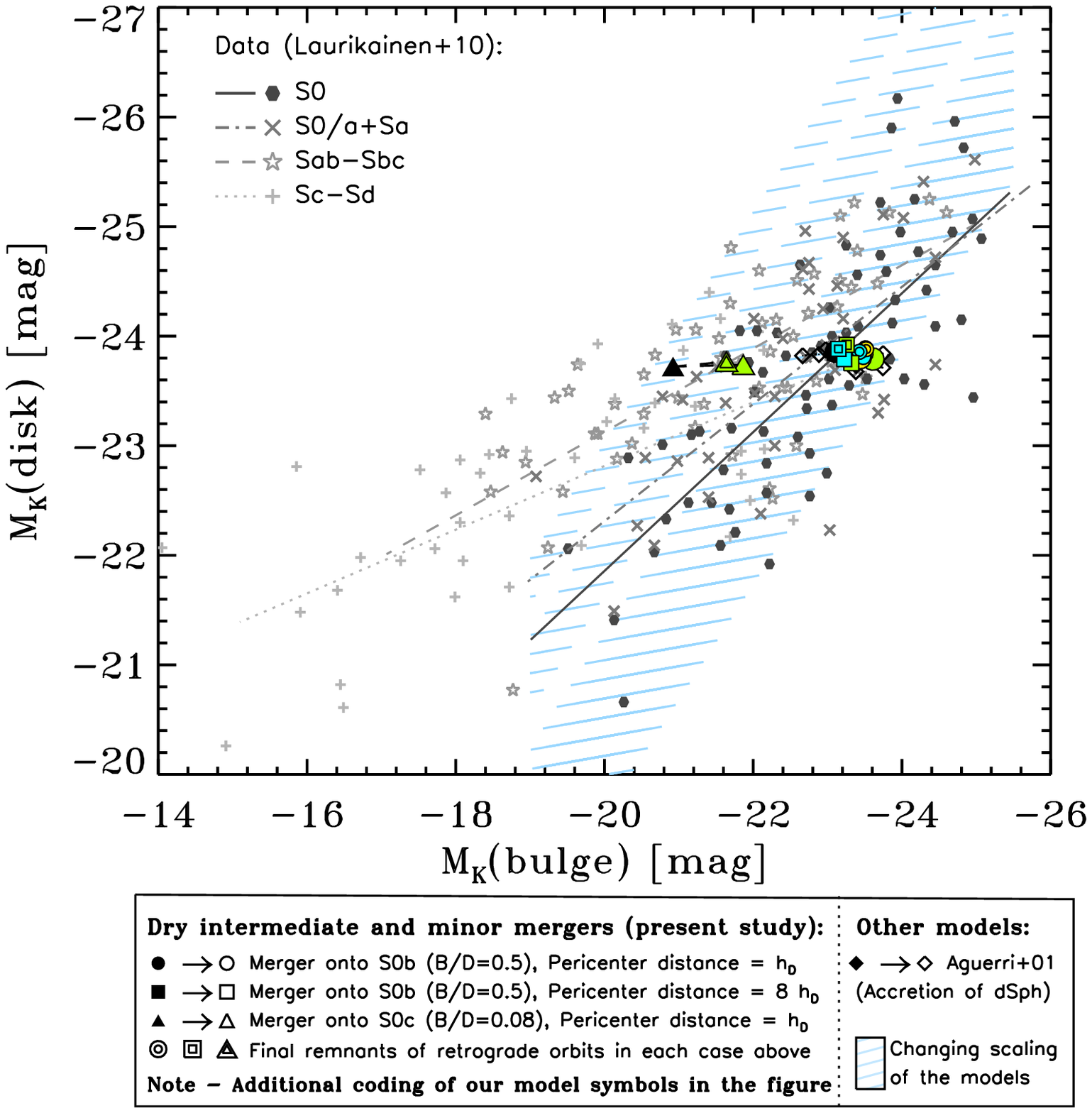} 
\caption{Evolution in the $M_K(\mathrm{disk})$-$r_\mathrm{eff}$ (\emph{left}) and $M_K(\mathrm{disk})$-$M_K(\mathrm{bulge})$ (\emph{right}) planes driven by our minor merger experiments, compared to the observational distributions of S0s and spirals by L10. The straight lines represent the best fits for S0s (solid lines), S0/a$+$Sa (dotted-dashed lines), and Sab-Sbc galaxies (dashed lines) obtained by L10 in each diagram, all with statistically significant rank correlation coefficients. The blue shaded region \emph{in the right panel} indicates the locations in the plane that are covered by our models with a different mass scaling for each location. Consult the legend for the observational data in the figure. Growth vectors have been plotted for the models, but the evolution induced by the mergers in the planes is negligible. The legend for the models is the same as in Fig.\,\ref{fig:kormendy}. }\label{fig:bulgedisk}
\end{figure*}

In Fig.~\ref{fig:disk2}, we represent the growth vectors induced by our merger models in the $\mu_\mathrm{0,D}$-$M_K(\mathrm{disk})$ plane. The mergers can modify the disk central surface brightness by up to $\sim 1$\,mag, which induces a negligible change in the disk magnitude. This contributes to disperse the remnants in the plane, showing a wide range of $\mu_\mathrm{0,D}$ values for a given value of the disk magnitude, in agreement with S0 data. This explains why the disks of S0s do not follow the Freeman law \citep{1970ApJ...160..811F}, as reported by L10 and references therein. 

We do not find clear trends with the mass ratio of the encounter, the central satellite density, or the spin-orbit coupling in the photometric relations between the disk parameters shown in Figs.~\ref{fig:disk} and \ref{fig:disk2}, which also occurred in the trends relating bulge parameters (see Sect.~\ref{Sec:bulgeall}). However, the changes induced by the mergers in these photometric planes exhibit clear trends depending on the orbital pericentre distance. Long-pericentre orbits tend to generate more concentrated disks than initially, whereas short-pericentre orbits tend to expand them. This is because the models with long pericentres decrease $h_\mathrm{D}$ and brighten $\mu_\mathrm{0,D}$, whereas the experiments with short-pericentre orbits do the opposite in general: they increase $h_\mathrm{D}$ and dim $\mu_\mathrm{0,D}$. This trend is exactly the opposite to the one found for the bulges (see Sect.~\ref{Sec:bulge}). Also in contrast to the trends involving bulge parameters, the changes in the disk parameters induced by mergers with short-pericentre orbits are not necessarily more noticeable than changes in long-pericenter orbits.

We find that the relations of the disk parameters in our remnants are similar to those of spirals, in agreement with S0 data (L10). We can therefore conclude that dry intermediate and minor mergers also agree with the photometric relations of disk parameters in S0s.

\subsection{Photometric relations of bulge and disk parameters}
\label{Sec:bulgedisk}

Figure~\ref{fig:bulgedisk} shows the evolution induced by our merger models in the planes relating the disk magnitude with other bulge parameters (in particular, with $r_\mathrm{eff}$ and its magnitude). L10 reported noticeable correlations in these photometric planes, similar again for all morphological types (except that the slope is slightly steeper for the S0s in the $M_K(\mathrm{disk})$-$M_K(\mathrm{bulge})$ relation). 

Our merger models induce negligible changes in both the disk magnitude and the bulge effective radius, which contributes to preserve any relation between these parameters that could exist prior to the merger (see the left panel of the figure). Therefore, as occurred in the $h_\mathrm{D}$-$M_K(\mathrm{disk})$ relation (see Fig.~\ref{fig:disk}), dry intermediate and minor mergers are compatible with the observational fact that the $r_\mathrm{eff}$-$M_K(\mathrm{disk})$ correlation is common to all morphological types, because they just introduce a slight dispersion in it.

The mergers also tend to slightly brighten the bulges, moving the remnants towards the right in the $M_K(\mathrm{disk})$-$M_K(\mathrm{bulge})$ plane, where S0s accumulate (see the right panel of Fig.~\ref{fig:bulgedisk}). Considering a different mass scaling for our models, we can basically reproduce the location of S0s at the right of this plane (see the dashed blue area marked in the diagram).  
 
Again, we do not find clear trends with the mass ratio of the encounter in the photometric relations shown in Fig.~\ref{fig:bulgedisk}, i.e., the changes in the photometric parameters are not necessarily more pronounced in the encounters with mass ratios 6:1 than in those with 18:1. This lack of dependence on the mass ratio seems to contradict previous findings \citep{2003ApJ...597..893N,2005A&A...437...69B}. But the initial conditions in these studies and ours are different because these authors started with spirals and not with S0s. Moreover, the mass dependence could be present if lower mass ratios are considered, as occurs in Fig.\,10 in \citet{2005A&A...437...69B}. We find no dependency on the spin-orbit coupling and the central satellite density, nor clear trends with the orbital pericentre distance (in contrast to the photometric scaling relations involving only bulge or disk parameters).

Summarising, dry intermediate and minor mergers onto S0s would preserve the bulge-disk coupling that could exist prior to the merger in the original galaxy regarding the $M_K(\mathrm{disk})$-$r_\mathrm{eff}$ and $M_K(\mathrm{disk})$-$M_K(\mathrm{bulge})$ relations, naturally contributing to the offset of S0s in this last relation, which agrees well with observations (L10).

\begin{sidewaysfigure*}[!]
\vspace{2cm}
\includegraphics[width=0.5\textwidth]{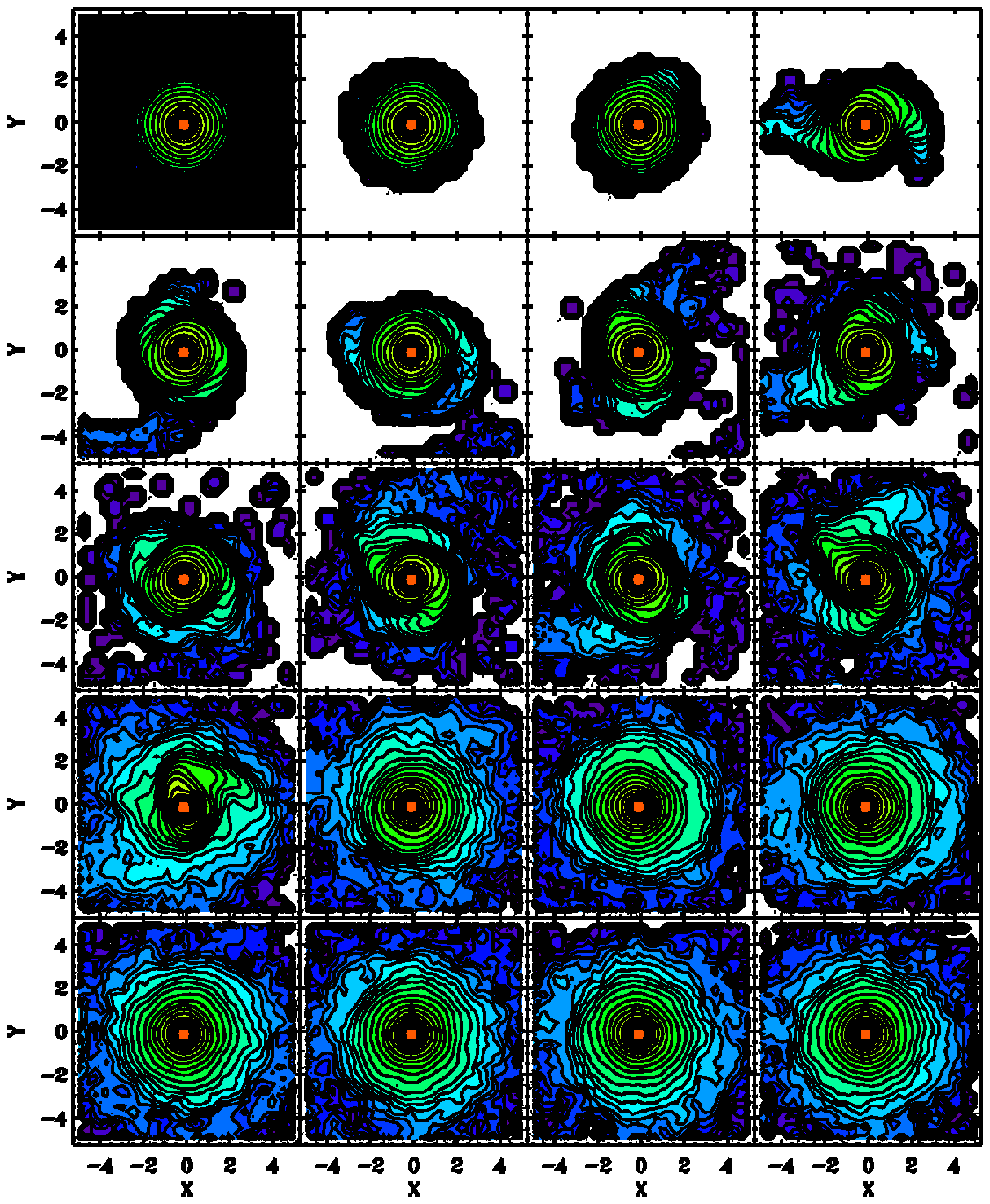}
\includegraphics[width=0.5\textwidth]{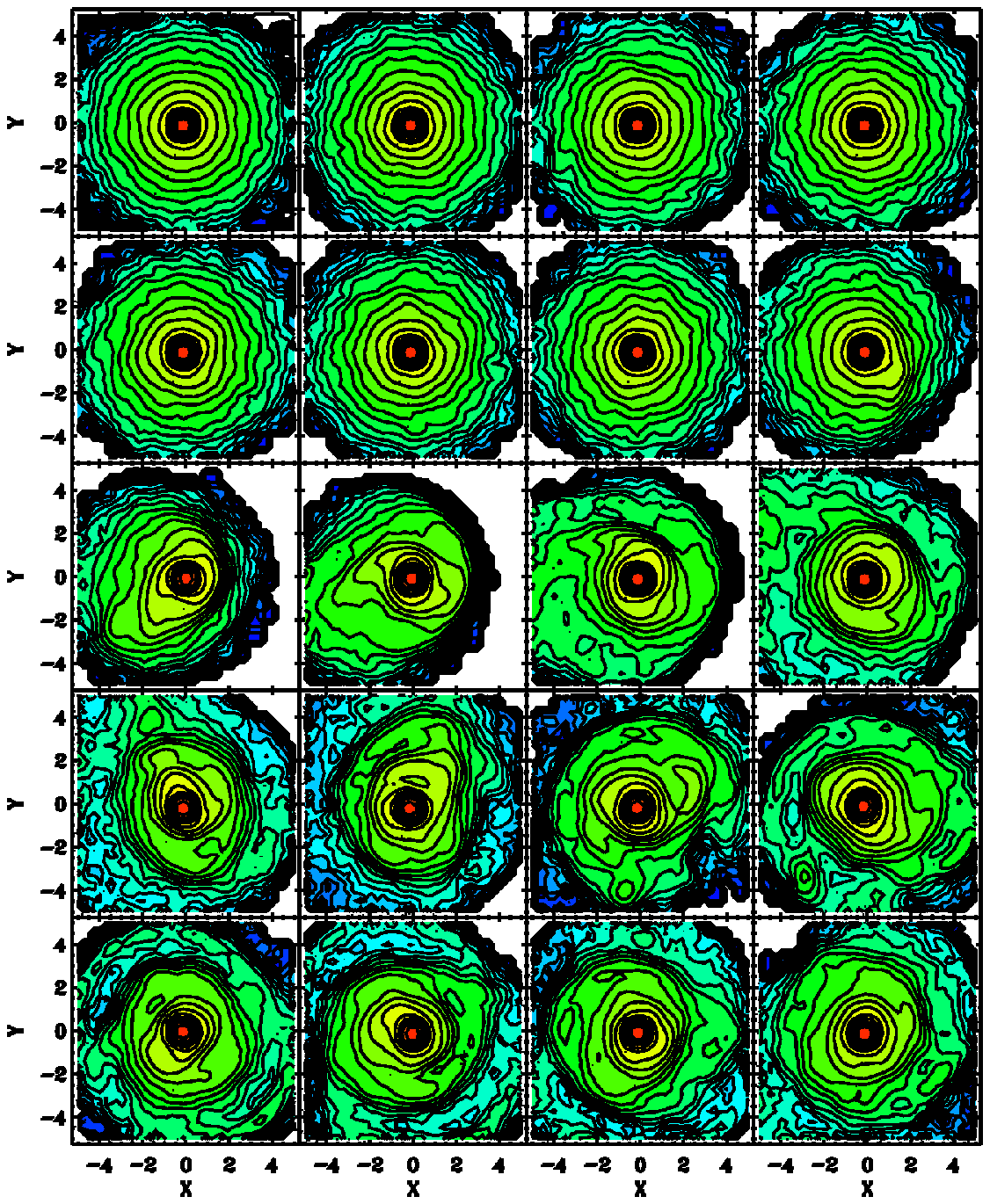}
\caption{Time evolution of the luminous surface density of the primary disk material in a direct model with a large pericentre distance and mass ratio 18:1 (model M18PlDb, left panels) and in a retrograde model with a short pericentre and lower mass ratio (model M9PsRb, right panels), both assuming an initial S0b galaxy. Face-on views centered on the initial primary galaxy are plotted. Snapshots corresponding to times from 20 to 260 and from 20 to 100 are shown from top-to-bottom and left-to-right of each figure, using a time step equal to 12 and to 4 in each case (in simulation units). Scaling the primary S0b galaxy in both models to the MW (see Sect.~\ref{Sec:scaling}), the total time period represented in each figure corresponds to $\sim 4.5$\,Gyr and $\sim 1.6$\,Gyr, respectively. A rainbow colour palette is used to represent different surface density levels in a logarithmic scale, with redder colours indicating higher values. Spatial scales in both axes are provided in simulation units.  
}\label{fig:spiral1}
\end{sidewaysfigure*}

\section{Discussion}
\label{Sec:discussion}

In this study and in Paper I, we have shown that dry intermediate and minor mergers onto S0s induce evolution within the sequence of S0 Hubble types, generating S0 remnants with structural and photometric properties that agree with observations. It is remarkable that, in particular, this is also true in the photometric scaling relations which point to a strong bulge-disk coupling. Therefore, the present models disprove the popular view that mergers unavoidably decouple the structure of these two components in galaxies.

We find that our simulated mergers induce evolution in the photometric planes in two ways. In some of the photometric relations, the encounters induce a negligible evolution with respect to the original location of the galaxy in it; they merely cause a slight dispersion compatible with observations. Therefore, if the original galaxies fulfill these relations prior to a merger, their remnants will obey them too, as occurs in the $M_K(\mathrm{disk})$-$r_\mathrm{eff}$ and $h_\mathrm{D}$-$r_\mathrm{eff}$ relations (see Fig.~\ref{fig:bulgedisk} here and Fig.~3 in Paper I). This means that, although these mergers do not build up these photometric relations, neither do they destroy them. In other cases, the mergers drive the remnants to build up the observed trend, as occurs in the $M_K(\mathrm{disk})$-$M_K(\mathrm{bulge})$ and $h_\mathrm{D}$-$n$ planes (see Fig.~\ref{fig:bulgedisk} here and Fig.~4 in Paper I). Independently of the case, the simulations show that mergers can preserve or even reinforce the tight structural binding between the bulge and disk components observed in S0s. 

There are two physical processes in these simulations responsible of the preservation of the bulge-disk coupling. The first mechanism is that the merger triggers an inward transport of primary disk material through disk instabilities, which contributes more to the growth of the bulge than the deposition of disrupted satellite material itself (EM06). In Fig.\,\ref{fig:spiral1}, we show the time evolution of the spiral patterns and oval distortions induced by the mergers for two models. The inflows of disk material towards the galaxy centre through these transient spirals and ovals (a type of evolution normally ascribed to pure internal secular evolution) contribute to link the bulge to its host disk both structurally and dynamically, even though the process has been triggered outside the galaxy.  \emph{In this sense, minor and intermediate-mass mergers induce internal secular evolution. }

The second mechanism is twofold. On one side, the density profile of the original disk partially survives the merger, curving up inwards and expanding outwards after the merger \citep[EM06, see also][]{2004A&A...418L..27B,2007ApJ...670..269Y}. On the other side, the satellite stars are located across a wide radial range in the remnants, rebuilding the exponential profile of the disk at intermediate radii \citep[EM06, but see also][]{2004A&A...418L..27B}. The accreted satellite stars end in fast-rotating orbits, building up thin rotationally supported inner components out of disrupted satellite material (EM11). This enhances the structural and dynamical coupling between bulge and disk in the remnant S0.

In all models with original S0b galaxies, the merger-induced spirals and distortions are efficiently diluted in a time period $\sim 0.5$-1\,Gyr after the full merger (see Fig.\,\ref{fig:spiral1}). This quick stabilization of the disk is caused by a central massive component in the original galaxy \citep[as noted previously by many authors, see][]{1990ApJ...363..391P,1999ApJ...510..125S,2002A&A...392...83B,2005MNRAS.363..496A,2005ApJ...628..678D,2005MNRAS.357..753G,2008MNRAS.384..386C}. The remnants of galaxies with initial small bulges (S0c) still exhibit significant spiral patterns and central oval distortions at the end of the simulation because they do not inhibit them as efficiently after the full merger (see Fig.~1 in Paper I).

Figure~\ref{fig:spiral1} also shows that mergers with short-pericentre orbits and high satellite masses generate more noticeable structural distortions in the primary disk than long-pericentre orbits or low-mass accretions (compare the two models shown in the figure). We have seen in this study that encounters with mass ratios 6:1 do not necessarily entail more noticeable changes in the photometric parameters than those with less massive satellites. In fact, we have not found any trends in the photometric relations as a function of the mass ratio, the spin-orbit coupling, or the satellite central density for the range of values studied here (but see Section\,\ref{Sec:bulgedisk}). The orbital pericentre distance seems to segregate the remnants in the photometric planes more clearly  (see Fig.~\ref{fig:nmkbulge}), sometimes even inducing opposite changes in the photometric parameters depending on it (as occurs in $\mu_\mathrm{0,D}$, $h_\mathrm{D}$, $\mu_\mathrm{eff}$, and $r_\mathrm{eff}$, see Table\,\ref{tab:photometricparameters}). 

As commented above, most remnants harbour small ovals and dynamically cold flat inner components consisting of disrupted satellite material (such as inner disks, inner rings, pseudo-rings, nested inner disks, and central spiral patterns), all phenomena frequently found in real S0s \citep{2005MNRAS.362.1319L,2006AJ....132.2634L,2009ApJ...692L..34L,2011MNRAS.418.1452L}. These components naturally establish a structural link between the bulge and the disk in the galaxy as well. Accordingly, their existence in S0s, usually ascribed to pure internal secular processes in galaxies, may ultimately be related to minor mergers. 

It is remarkable that the mergers preserve or even reinforce the bulge-disk coupling in the resulting galaxies, even though no remnant develops a bar. As commented above, our primary disks are resistant to instabilities (and in particular, to bars) because of large mass concentrations in the original galaxies: the initial S0b's have a prominent bulge ($B/D=0.5$), while the original S0c's require concentrated dark matter haloes to have stable disks. This fact, instead of being a limitation of the models, should be considered as a strong support to the idea that most minor mergers induce significant internal secular evolution, even in the absence of bars. Indeed, our bulges increase their masses by $\sim 20$-50\% of their original masses, according to the photometric decompositions (see Table\,\ref{tab:photometricparameters}). 

Nevertheless, we must note that we used gas-free simulations. S0s do contain less gas than spirals, but they still contain some cold (HI and/or H$_\mathrm{2}$) gas \citep{2011MNRAS.410.1197C,2011MNRAS.415...32S}. This small gas amount could affect the outcome of the minor mergers simulated in this paper to some degree. If the effects of dissipation and star formation are taken into account, the central parts of the remnants should probably become bluer and denser, with the bulges growing even larger. In any case, the small amount of gas in S0s probably prevents these effects from becoming larger than the bulge mass growth induced by the merger itself, which can be as high as $\sim 50$\% of its original mass, as noted above. We would require gas amounts too high for the mass ratios studied here to induce star formation rates capable of increasing the bulge mass by a similar fraction \citep{2008MNRAS.384..386C,2011ARep...55..770T}. Moreover, the evolution simulated here must become relevant in groups, as noted in Section\,\ref{Sec:introduction}, and many groups are located in dense environments (clusters), where gas is stripped \citep{2000Sci...288.1617Q}. In particular, the NIRS0S sample we used as a comparison throughout the paper is magnitude-limited \citep{2011MNRAS.418.1452L} and includes galaxies in all kinds of environments. However, because the galaxies are nearby systems, the only cluster in the sample is Virgo, to which 28 galaxies of the sample belong (E.~Laurikainen, private communication). Accounting for the fact that $\sim 60$\% of S0s reside in groups (see references in Section~\ref{Sec:introduction}), the average galaxies of the comparison sample might basically belong to groups, an environment where mergers can be significant evolutionary mechanisms. 

All these arguments indicate that internal and external secular evolutionary processes may be strongly related, in the sense that the latter usually trigger the former ones. Our models show that minor mergers can induce significant internal secular evolution even without bars and dissipational effects \citep[see also][]{2003MNRAS.338..880S}. Therefore, it must be difficult to isolate the contribution to the bulge growth of pure internal secular evolution from the contribution of merger-driven internal secular processes in present-day galaxies. Recent observations and simulations seem to confirm this \citep{2009A&A...496...51P,2012MNRAS.427.1503Z,2013MNRAS.429.1051C}. Moreover, our models also refute the popular view that mergers cannot conserve the bulge-disk coupling observed in the photometric scaling relations of S0s. In fact, they prove the capability of these evolutionary mechanisms to preserve (and even strengthen) the structural bulge-disk coupling in the galaxies that undergo them.

Our results show that intermediate and minor mergers induce a smooth evolution in the original S0 galaxy, at least regarding their photometric scaling relations. Considering the huge relevance of mergers and tidal interactions for galaxy evolution in some environments \citep[as galaxy groups and even clusters, see][]{2007ARep...51..435T,2012MNRAS.423.1277D}, these processes must have contributed noticeably to the properties of present-day S0s.

\section{Conclusions} 
\label{Sec:conclusions}

Using N-body simulations, we have studied whether the evolution induced in S0s by dry intermediate and minor mergers can reproduce the photometric scaling relations of these galaxies or not, particularly those pointing to a strong bulge-disk coupling. Our models show that these mergers can induce internal secular evolution that generates S0 remnants that fulfill the photometric scaling relations observed for these galaxies. It is remarkable that this is also true for the scaling relations pointing to a strong bulge-disk coupling in S0s, which disproves the common view that mergers unavoidably decouple the structure of these two galactic components.
 
Mergers affect the photometric relations in two ways. They move the remnants along the observed photometric relation in some cases (e.g., the photometric plane of the bulges or the $\mu_\mathrm{0,D}$-$h_\mathrm{D}$ relation), while they just induce a slight dispersion compatible with data in others  (e.g., the relations involving the disk magnitude). In the photometric planes where the morphological types tend to segregate (as occurs in most relations involving the parameters of the bulge), the mergers always induce evolution towards the region populated by S0s in a way compatible with observations.

We find no clear trends with the mass ratio, the satellite central density, and the spin-orbit coupling of the encounter for the range of values studied here. However, we report clear trends with the orbital pericentre orbit: long-pericentre orbits generate more concentrated disks and more expanded bulges than initially, whereas orbits with short pericentre distances tend to expand disks and compress bulges. We confirm that the evolution triggered in the models with a primary S0c galaxy is more noticeable than in those with an S0b, as already noted in paper I.

The structural bulge-disk coupling is preserved or reinforced in these models by two facts: first, because the merger does not completely destroy the original galaxy structure, and secondly, because the interaction triggers some internal secular processes in the primary disk that entail bulge growth (transient ovals and spirals). It is remarkable that no significant bar is formed in any experiments.

Therefore, gas-poor mergers of low-mass ratios can be considered as plausible mechanisms for the evolution of S0s considering their photometric scaling relations because these models show that they tend to conserve (and even give rise to) these relations. The mergers can induce significant internal secular evolution that contributes to preserve or even reinforce the bulge-disk coupling in the remnant, even without bars or dissipational effects. Satellite accretions thus seem to unavoidably entail internal secular evolution, meaning that it may be quite complex to isolate the effects of the internal secular evolution driven by mergers from the one due to purely intrinsic disk instabilities in individual early-type disks at the present.

%
\small  
%
\begin{acknowledgements}   

The authors are very grateful to the referee, Frederic Bournaud, for the input that helped to improve this publication significantly. We also thank Olga Sil'chenko for stimulating the present study, as well as Miguel Querejeta and Eija Laurikainen for interesting and useful comments. Supported by the Spanish Ministry of Science and Innovation (MICINN) under projects AYA2009-10368, AYA2006-12955, AYA2010-21887-C04-04, AYA2009-11137, and AYA2012-30717, and by the Madrid Regional Government through the AstroMadrid Project (CAM S2009/ESP-1496, http://www.laeff.cab.inta-csic.es/projects/astromadrid/main/index.php). Funded by the Spanish MICINN under the Consolider-Ingenio 2010 Program grant CSD2006-00070: "First Science with the GTC" (http://www.iac.es/consolider-ingenio-gtc/), and by the Spanish programme of International Campus of Excellence Moncloa (CEI). ACGG is a Ramon y Cajal Fellow of the Spanish MICINN. This research is based in part on services provided by the GAVO data center. It has made use of the NASA/IPAC Extragalactic Database (NED) which is operated by the Jet Propulsion Laboratory, California Institute of Technology, under contract with the National Aeronautics and Space Administration. 

\end{acknowledgements}

\begin{sidewaystable*}[!t]
\centering 
\caption{Bulge and disk photometric parameters of the original primary galaxies and of the final remnants in the merger experiments${^\mathrm{a,b,c}}$}\vspace{0.1cm}
\label{tab:photometricparameters}
{\small
\centering
\begin{tabular}{lccccccccc}
\hline\hline   
\multicolumn{1}{c}{Model} & \multicolumn{1}{c}{$\mu_\mathrm{eff}$} & \multicolumn{1}{c}{$r_\mathrm{eff}$} &  \multicolumn{1}{c}{$n$} &  \multicolumn{1}{c}{$\mu_\mathrm{0,D}$} &  \multicolumn{1}{c}{$h_\mathrm{D}$} &  \multicolumn{1}{c}{$B/D$} &  \multicolumn{1}{c}{$M_K(\mathrm{bulge})$} &  \multicolumn{1}{c}{$M_K(\mathrm{disk})$} &  \multicolumn{1}{c}{$M_K(\mathrm{total})$}\\
\multicolumn{1}{c}{}  &  \multicolumn{1}{c}{[mag/arcsec$^2$]} & \multicolumn{1}{c}{[Kpc]} &  \multicolumn{1}{c}{} & 
\multicolumn{1}{c}{[mag/arcsec$^2$]} & \multicolumn{1}{c}{[Kpc]} & \multicolumn{1}{c}{} & \multicolumn{1}{c}{[mag]} & \multicolumn{1}{c}{[mag]} & \multicolumn{1}{c}{[mag]}\\
\multicolumn{1}{c}{(1)} & \multicolumn{1}{c}{(2)} & \multicolumn{1}{c}{(3)} & \multicolumn{1}{c}{(4)}  & \multicolumn{1}{c}{(5)} & \multicolumn{1}{c}{(6)} & \multicolumn{1}{c}{(7)} & \multicolumn{1}{c}{(8)} & \multicolumn{1}{c}{(9)} & \multicolumn{1}{c}{(10)}\\\hline\vspace{-0.3cm}\\
 Original S0b ($B/D=0.5$) & 14.90$\pm$0.08  &   0.90$\pm$0.04 & 0.92$\pm$0.11 & 17.10$\pm$0.07 & 4.73$\pm$0.10 & 0.51$\pm$0.03 & -23.11$\pm$0.24 & -23.86$\pm$0.11 & -24.30$\pm$0.12\\
 Original S0c ($B/D=0.08$) & 15.97$\pm$0.07 & 0.428$\pm$0.007 & 0.52$\pm$0.02 & 16.71$\pm$0.07 & 2.66$\pm$0.03 & 0.0760$\pm$0.0024 & -20.92$\pm$0.12 & -23.72$\pm$0.09 & -23.80$\pm$0.10   \vspace{0.05cm}\\\hline\vspace{-0.3cm}\\
  (a) M6 Ps Db  & 14.52$\pm$0.09 & 0.82$\pm$0.06 & 1.7$\pm$0.4 & 17.25$\pm$0.07 & 5.00$\pm$0.23 & 0.82$\pm$0.10 & -23.6$\pm$0.4 & -23.83$\pm$0.17 & -24.47$\pm$0.18\\
  (a2) M6 Ps Db TF3  & 14.65$\pm$0.09 & 0.86$\pm$0.06 & 1.6$\pm$0.3 & 17.35$\pm$0.07 & 5.22$\pm$0.18 & 0.77$\pm$0.08 & -23.5$\pm$0.4 & -23.82$\pm$0.14 & -24.45$\pm$0.16\\
  (a3) M6 Ps Db TF4  & 14.62$\pm$0.10 & 0.86$\pm$0.10 & 1.8$\pm$0.6 & 17.32$\pm$0.07 & 5.09$\pm$0.27 & 0.9$\pm$0.3 & -23.6$\pm$0.6 & -23.79$\pm$0.19 & -24.46$\pm$0.22\\
  (b) M6 Ps Rb &  14.65$\pm$0.09 & 0.84$\pm$0.06 & 1.7$\pm$0.4 & 17.02$\pm$0.07 & 4.64$\pm$0.14 & 0.70$\pm$0.11 & -23.5$\pm$0.4 & -23.89$\pm$0.13 & -24.47$\pm$0.15\\
 (c) M6 Pl Db &  14.98$\pm$0.08 & 0.985$\pm$0.010 & 1.11$\pm$0.04 & 16.90$\pm$0.07 & 4.11$\pm$0.04 & 0.676$\pm$0.024 & -23.32$\pm$0.12 & -23.75$\pm$0.09 & -24.31$\pm$0.09\\
 (d) M6 Pl Rb  & 14.96$\pm$0.08 & 0.97$\pm$0.014 & 1.02$\pm$0.06 & 16.40$\pm$0.07 & 3.53$\pm$0.03 & 0.55$\pm$0.03 & -23.26$\pm$0.14 & -23.92$\pm$0.09 & -24.39$\pm$0.10\\
 (e) M6 Ps Ds &  15.78$\pm$0.08 & 0.44$\pm$0.03 & 2.29$\pm$0.16 & 16.78$\pm$0.07 & 2.77$\pm$0.07 & 0.179$\pm$0.012 & -21.9$\pm$0.3 & -23.74$\pm$0.13 & -23.92$\pm$0.14\\
 (f) M6 Ps Rs  & 15.55$\pm$0.08 & 0.35$\pm$0.03 & 2.44$\pm$0.22 & 16.51$\pm$0.07 & 2.48$\pm$0.06 & 0.140$\pm$0.012 & -21.6$\pm$0.4 & -23.77$\pm$0.12 & -23.91$\pm$0.14 \vspace{0.05cm}\\\hline\vspace{-0.3cm}\\
 (g) M9 Ps Db  & 14.52$\pm$0.09 & 0.82$\pm$0.05 & 1.5$\pm$0.3 & 17.30$\pm$0.07 & 5.13$\pm$0.18 & 0.77$\pm$0.08 & -23.5$\pm$0.4 & -23.83$\pm$0.14 & -24.44$\pm$0.16\\
(g2) M9 Ps Db TF3  & 14.60$\pm$0.09 & 0.83$\pm$0.07 & 1.3$\pm$0.3 & 17.22$\pm$0.07 & 5.00$\pm$0.14 & 0.65$\pm$0.06 & -23.4$\pm$0.4 & -23.85$\pm$0.13 & -24.40$\pm$0.14\\
(g3) M9 Ps Db TF4  & 14.75$\pm$0.08 & 0.90$\pm$0.05 & 1.3$\pm$0.3 & 17.35$\pm$0.07 & 5.18$\pm$0.18 & 0.75$\pm$0.08 & -23.4$\pm$0.4 & -23.80$\pm$0.14 & -24.39$\pm$0.16\\
(h) M9 Ps Rb  & 14.52$\pm$0.11 & 0.78$\pm$0.11 & 1.9$\pm$0.8 & 17.07$\pm$0.07 & 4.73$\pm$0.23 & 0.72$\pm$0.18 & -23.5$\pm$0.7 & -23.88$\pm$0.17 & -24.47$\pm$0.21 \vspace{0.05cm}\\\hline\vspace{-0.3cm}\\
(i) M18 Ps Db  & 14.52$\pm$0.11 & 0.81$\pm$0.10 & 1.5$\pm$0.3 & 17.32$\pm$0.07 & 5.13$\pm$0.23 & 0.60$\pm$0.08 & -23.5$\pm$0.5 & -23.81$\pm$0.16 & -24.41$\pm$0.18\\
(j) M18 Ps Rb  & 14.55$\pm$0.09 & 0.79$\pm$0.07 & 1.6$\pm$0.3 & 17.20$\pm$0.07 & 4.95$\pm$0.14 & 0.68$\pm$0.07 & -23.4$\pm$0.4 & -23.86$\pm$0.13 & -24.47$\pm$0.14\\
(k) M18 Pl Db  & 14.96$\pm$0.08 & 0.945$\pm$0.015 & 1.05$\pm$0.05 & 16.95$\pm$0.07 & 4.28$\pm$0.07 & 0.60$\pm$0.03 & -23.22$\pm$0.14 & -23.79$\pm$0.10 & -24.30$\pm$0.10\\
(l) M18 Pl Rb  & 14.92$\pm$0.08 & 0.896$\pm$0.014 & 1.06$\pm$0.05 & 16.98$\pm$0.07 & 4.54$\pm$0.06 & 0.51$\pm$0.03 & -23.15$\pm$0.14 & -23.88$\pm$0.10 & -24.33$\pm$0.10  \\\hline\\

\end{tabular}
\tablefoot{\emph{Columns}: (1) Model code from Table\,\ref{tab:models}; (2) effective surface brightness of the bulge in mag/arcsec$^2$; (3) bulge effective radius in Kpc; (4) bulge Sersic index; (5) central surface brightness of the disk in mag/arcsec$^2$; (6) disk scale-length in Kpc; (7) bulge-to-disk luminosity ratio in the $K$ band; (8) $K$-band absolute magnitude of the bulge; (9) $K$-band absolute magnitude of the disk; (10) $K$-band total absolute magnitude. \tablefoottext{a}{The photometric parameters are provided for the $K$ band. A mass-to-light ratio in the $K$ band equal to $M/L_K=1$ is assumed to transform surface density profiles in the models to surface brightness profiles (see Sect.~\ref{Sec:scaling}).} \tablefoottext{b}{The values listed in this table assume the scaling discussed in Sect.~\ref{Sec:scaling}, i.e., the original primary galaxy with $B/D=0.5$ is scaled to the mass and size of the Milky Way, and the one with $B/D=0.08$ is scaled to NGC\,253. We have kept these scalings throughout the paper to facilitate the comparison with observational data, but they can be modified by assuming different mass and size units.} \tablefoottext{c}{Magnitudes are in the Vega system.}
}
}
\end{sidewaystable*}

\bibliographystyle{aa}
\bibliography{elic0709_def.bib}{}
\end{document}